%% file: main_arxiv_submission.tex
\documentclass{article}

\usepackage{microtype}
\usepackage{graphicx}
\usepackage{subfigure}
\usepackage{booktabs} %
\usepackage{siunitx}

\usepackage{hyperref}

\usepackage[accepted]{preprint}

\usepackage{amsmath}
\usepackage{amssymb}
\usepackage{mathtools}
\usepackage{amsthm}

\usepackage[capitalize,noabbrev]{cleveref}
\DeclareMathOperator*{\argmax}{arg\,max}

\theoremstyle{plain}

\theoremstyle{definition}

\theoremstyle{remark}

\usepackage[textsize=tiny]{todonotes}

\input{icml2024/commands}
\icmltitlerunning{Weather Prediction with Diffusion Guided by Realistic Forecast Processes}

\begin{document}

\twocolumn[
\icmltitle{Weather Prediction with Diffusion Guided by Realistic Forecast Processes}

\icmlsetsymbol{equal}{*}

\begin{icmlauthorlist}
\icmlauthor{Zhanxiang Hua}{equal,uw-atms}
\icmlauthor{Yutong He}{equal,yyy}
\icmlauthor{Chengqian Ma}{uw-amath}
\icmlauthor{Alexandra Anderson-Frey}{uw-atms}
\end{icmlauthorlist}
\icmlaffiliation{uw-atms}{Department of Atmospheric Sciences, University of Washington, USA}
\icmlaffiliation{uw-amath}{Department of Applied Mathematics, University of Washington, USA}
\icmlaffiliation{yyy}{Machine Learning Department, Carnegie Mellon University, USA}

\icmlcorrespondingauthor{Zhanxiang Hua}{zxhua@uw.edu}

\vskip 0.3in
]

\printAffiliationsAndNotice{\icmlEqualContribution} 

\input{icml2024/tex/abstract_v2}
\input{icml2024/tex/intro_v2}

\input{icml2024/tex/related_works_v2}
\input{icml2024/tex/method}

\input{icml2024/tex/experiment}
\input{icml2024/tex/conclusion}
\newpage
\input{icml2024/tex/impact_statement}

\input{icml2024/tex/acknowledge}

\bibliography{example_paper}
\bibliographystyle{icml2024}

\newpage
\appendix
\onecolumn
\input{icml2024/tex/appendix}

\end{document}

%% file: icml2024/tex/abstract_v2.tex
\begin{abstract}
Weather forecasting remains a crucial yet challenging domain, where recently developed models based on deep learning (DL) have approached the performance of traditional numerical weather prediction (NWP) models. However, these DL models, often complex and resource-intensive, face limitations in flexibility post-training and in incorporating NWP predictions, leading to reliability concerns due to potential unphysical predictions. In response, we introduce a novel method that applies diffusion models (DM) for weather forecasting. In particular, our method can achieve both direct and iterative forecasting with the same modeling framework. Our model is not only capable of generating forecasts independently but also uniquely allows for the integration of NWP predictions, even with varying lead times, during its sampling process. The flexibility and controllability of our model empowers a more trustworthy DL system for the general weather community. Additionally, incorporating persistence and climatology data further enhances our model’s long-term forecasting stability. Our empirical findings demonstrate the feasibility and generalizability of this approach, suggesting a promising direction for future, more sophisticated diffusion models without the need for retraining.
\end{abstract}

%% file: icml2024/tex/intro_v2.tex
\section{Introduction}
\begin{figure}[ht]
\begin{center}
\centerline{\includegraphics[width=\columnwidth]{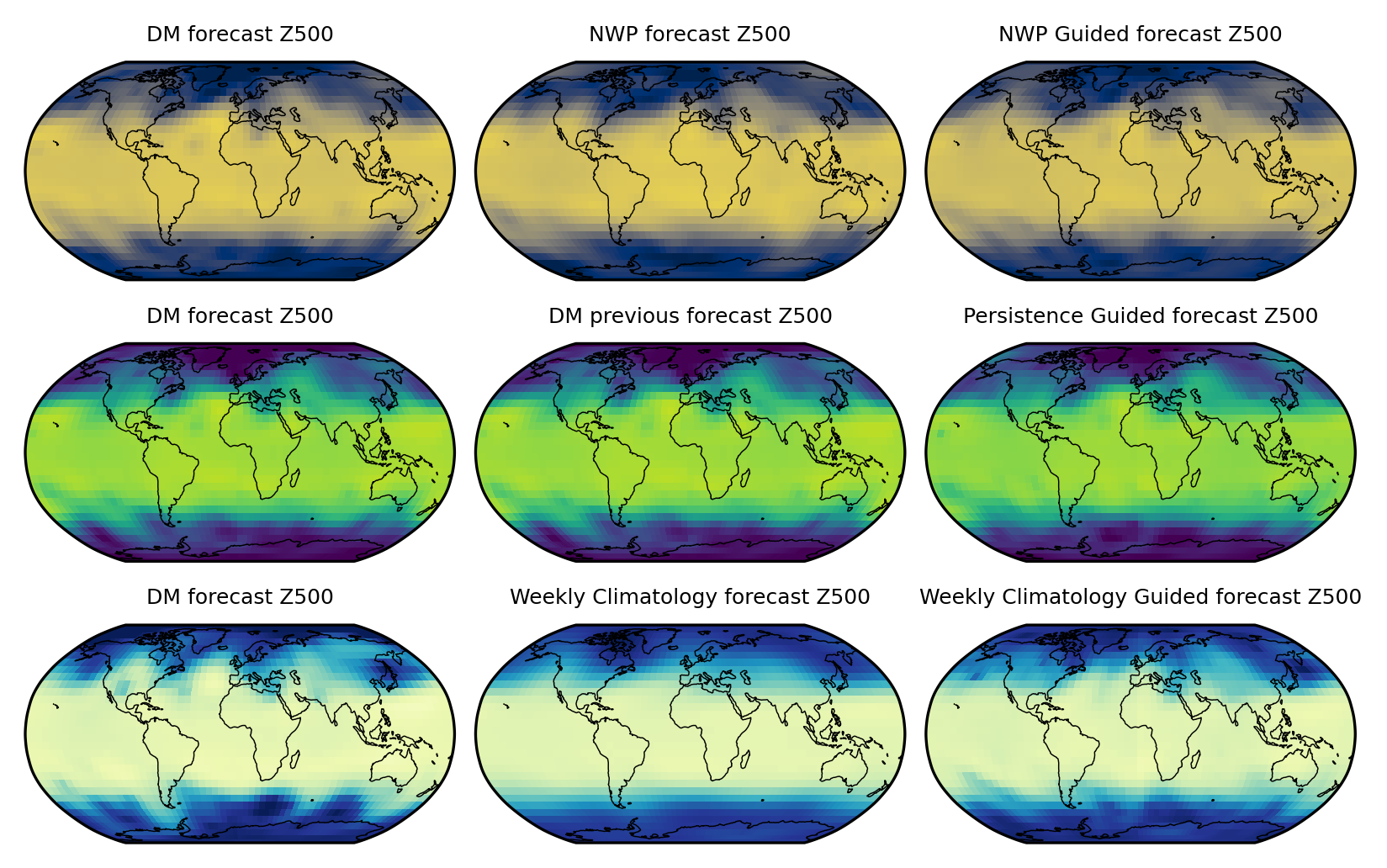}}
\caption{Illustration of the proposed three types of guidance that are relevant to weather forecasting using diffusion models. Our model is capable of incorporating all three types of guidance to facilitate more realistic weather forecasting with diffusion models.}
\label{intro}
\end{center}
\vskip -0.2in
\end{figure}

Weather forecasting, a key element of scientific computing, is essential for predicting future weather, especially extreme conditions like floods, droughts, and hurricanes. This forecasting is crucial across various sectors, impacting daily life, agriculture, energy, transportation, and industry \cite{bouwer2019observed}. Recent advancements in deep learning (DL) have led to data-driven models that rival traditional Numerical Weather Prediction (NWP) models \cite{bi2023accurate}. 
While data-driven DL models are gaining traction, the complexity and chaos of atmospheric systems pose significant challenges to these models, often leading to difficulties in accurately capturing the nonlinear nature of weather patterns. In addition, most state-of-the-art DL models are exceedingly complex, making them expensive and difficult for low-resource users to train.
More often, DL models are trained for weather prediction at a specific future time, requiring the retraining of multiple models if the existing autoregressive forecast falls short of the desired forecast period. Last but not least, once the model training is completed, it is extremely difficult to modify the model parameters or incorporate additional constraints such as physical guidance from NWP models at inference time, causing unphysical hallucinations and unreasonable predictions that do not obey natural physical laws \cite{selz2023can}. As a result, even though these DL models have achieved superior performance in certain benchmarks \cite{rasp2023weatherbench}, their reliability in real-world forecasting workflows remains questionable, particularly in long-range forecasts \cite{chattopadhyay2023long}, contributing to trust issues in these weather models \cite{mcgovern2024developing}. 

Diffusion models (DM) have gained popularity in recent years as AI-generated art platforms like Midjourney and OpenAI’s DALL-E have grown in popularity~\citep{dalle}.
Built with flexible probabilistic inference properties, DMs not only showcase high quality samples but also demonstrate remarkable controllability that does not require retraining of the model~\citep{dps,freedom,mpgd}. 
As the image generation domain witnesses tremendous development of DM applications, however, there has been limited research utilizing DMs for weather forecasting. 
While recent work from Deepmind \cite{price2023gencast} showcases DM's potential in 
ensemble medium-range weather forecasting that surpasses traditional methods, little work has been shown to explore the unique flexibility and controllability of DMs that can facilitate more accurate and stable predictions in low resource settings.
Moreover, the current literature only views DMs as another black-box proxy for better prediction, but misses out on the opportunity to establish formal connections between the probabilistic inference in diffusion models and the generative modeling view of traditional weather forecasting.

In this paper, we introduce a new weather forecasting model using a conditional DM, offering the capability to flexibly output weather predictions for various future time points.
The conditional DM forecasting model enables both iterative and direct forecasting methods within a single framework, where iterative forecasts build upon each step's previous prediction, and direct forecasts generate specific future predictions from the initial input. In addition, we design the conditional diffusion model that explicitly takes previous weather observations, non-weather constant variables and the target lead time as inputs, and treat the target lead time as ``class labels''. By incorporating classifier-free guidance~\citep{ho2022classifier}, our model is able to gain benefit from learning both the prior and the posterior distribution of a weather state given the lead time.

To tackle hallucination issues and the long-term stability of data-driven models, our diffusion framework can also combine various forecasting processes without the need for retraining. 
More specificalluy, we enhance this conditional DM forecasting model by integrating SDEdit \cite{meng2021sdedit}, a tool that allows adjustable noise levels and input guidances, tailored to the user's trust in the DM and the guidance. We showcase the versatility of this feature in three applications: NWP guidance, persistence guidance, and climatology guidance, each providing unique solutions.
This feature is particularly useful in addressing the limitations of data-driven DL models, especially in generating long-term and realistic weather patterns. The flexibility provided by our framework also allows users to choose the most suitable forecasting method for their needs and establishes a unique balance between trust and realism, distinguishing it from other applications of DL models.
Furthermore, the framework opens up the potential integration of long-range forecast model outputs as guidance coupling with DMs primarily designed for short-term forecasting. This integration aims to produce forecasts with higher temporal resolution while ensuring a seamless transition in weather states, eliminating the discontinuity typically encountered when reinitializing short-term forecast models.

%% file: icml2024/tex/related_works_v2.tex
\section{Related Works}

In the domain of weather forecasting, there has been a paradigm shift from traditional numerical models, which used physical computer models to numerically solve atmospheric and ocean equations (as summarized by \cite{bauer2015quiet}), to deep learning. Recent years have seen the rise of DL in weather forecasting, with DL-based models now surpassing traditional NWP models, a trend highlighted by \cite{rasp2020weatherbench} and \cite{rasp2023weatherbench}. Notable advancements include DL models like KeislerNet \citep{keisler2022forecasting} challenging state-of-the-art NWP models such as the European Centre for Medium-range Weather Forecasts's (ECMWF) high-resolution forecast (HRES). Others, like FourCastNet \citep{pathak2022fourcastnet}, Pangu-Weather \citep{bi2023accurate}, and a series of new models including GraphCast \citep{lam2023learning}, FuXi \citep{chen2023fuxi}, FengWu \citep{chen2023fengwu}, and Stormer \citep{nguyen2023scaling}, have shown competitive or superior performance, indicating a significant advancement in the field.

Recent advancements in DL models have been geared towards probabilistic forecasting to better handle the unpredictability of weather and the butterfly effect (\citealt{lorenz1996essence}), leading to a shift from deterministic to stochastic ensemble forecast systems. These systems, discussed in studies like \citealt{palmer2019ecmwf}, aim to represent a broader potential variation of weather forecasts but struggle with extremely high computational costs. To overcome these issues, diffusion models have been increasingly utilized in probabilistic forecasting, showing effectiveness in precipitation nowcasting (\citealt{asperti2023precipitation}, \citealt{gao2023prediff}) and turbulent flow simulation (\citealt{kohl2023turbulent}, \citealt{cachay2023dyffusion}). A significant advancement in this field is the introduction of GenCast \citep{price2023gencast}, a diffusion-based ensemble model that has outperformed ECMWF's ensemble forecasting in most evaluation metrics.

However, the application of Diffusion Models (DMs) in weather forecasting is underexplored. For example, techniques like SDEdit introduce guidance and controllability that balances the preservation of input structure guidance with the production of realistic generated outputs. In this paper, we use this approach to enhance DMs in weather forecasting without retraining, overcoming traditional DL models' limitations by providing realistic long-term forecasts and outputs adjustable to user trust levels, and thereby advancing the stability and reliability of weather prediction models.

%% file: icml2024/tex/method.tex
\section{Method}
\subsection{Problem Statement}

The goal of this paper is to forecast future weather based on observed weather. The weather, or more formally weather state, $x \in \mathcal{X} \subset \mathbb{R}^{H\times W\times C_1}$ denotes a collection of variables that describe the atmosphere conditions at different pressure levels for a given time such as temperature, humidity, pressure field height, wind speed, etc. 
Each weather state is represented by a two-dimensional grid of $H$ latitude and $W$ longitude grid points with $C_1$ channels that represent different variables at different pressure levels. Setting the current time to be $0$, to represent how far into the future this prediction will extend, we denote the difference between the target future time and current time as lead time $K \in \mathbb{R}^+$. The target future weather state $x$ with a lead time $K$ is denoted as $x^{(K)} \in \mathcal{X}$.

We assume there exist $C_0$ non-weather constant conditions such as terrain height and land-sea mask for all locations on the $H\times W$ grid and $L$ observed current or prior weather states available. We then represent the non-weather constant conditions as $y \in \mathcal{Y} \subset \mathbb{R}^{H\times W\times C_0}$, and the $L$ available prior weather states as a spatial-temporal sequence $\Vec{x}_{\text{ob}} = \left[ x^{(k_1)}, x^{(k_2)}, \cdots, x^{(k_L)}\right] \in \mathcal{X}^L \subset \mathbb{R}^{L \times H \times W \times C_1}, k_l\in\mathbb{R}$.

With the context provided, we define our weather forecasting task as the following: Given a sequence of observed weather states $\Vec{x}_{\text{ob}}$ and constant variables $y$, what are the future weather states $x^{(K)}$ at lead time $K$? In terms of probabilistic modeling, we would like to obtain
\begin{equation}
    x^{(K)*} = \argmax_{x^{(K)}} P(x^{(K)}|y, x^{(k_1)}, x^{(k_2)}, \cdots, x^{(k_L)})
    \label{direct_obj}
\end{equation}

\subsection{Unified Framework}
To answer this question, there exist two major forecasting frameworks -- ``\textbf{direct}'' and ``\textbf{iterative}''.
The direct forecast aims for a mapping $f:\mathcal{Y}\times\mathcal{X}^L \to \mathcal{X}$ that directly outputs the predicted future weather state conditioned on all prior observations and conditions in one step following Eq~\eqref{direct_obj}.

The iterative forecast is similar to traditional NWP modeling: the forecast is generated recurrently by using the intermediate forecast output as part of the input conditions for the next step of the forecast until the lead time $K$ is reached.
Specifically, suppose $\Vec{x}_{\text{ob}} = [x^{(K_0)}, x^{(K_0 + \Delta k)}, ..., x^{(K_0 + (L-1)\Delta k)}]$ where $K_0$ is the earliest observed timestamp and $\Delta k$ is the lead time between each observed timestamp, we can decompose the direct objective into the product of intermediate steps by chain rule and perform autoregressive sampling with objective:
\begin{equation}
\small
    x^{(K_0 + N\Delta k)*} = \argmax_{x^{(K_0 + N\Delta k)}} \prod_{n=L}^N P(x^{(K_0 + n\Delta k)}|y,[x^{(K_0 + j\Delta k)}]_{j<n})
    \label{iterative_ar}
\end{equation}
For all $j \geq L$, $x^{(K_0 + j\Delta k)}$ is calculated as $x^{(K_0+j\Delta k)*}$ using Eq~\eqref{iterative_ar}. When $N = (K-K_0)/\Delta k$, we can obtain the target prediction at lead time $K$.

While direct models offer more advantages in parametrization flexibility and speed,
the iterative models allow forecasting weather states beyond the predefined lead time during training and thus allow the incorporation of guidance at any instance of lead time. Hence we ask the question: Is there a way to unify direct and iterative modeling to obtain the advantages from both types of models?

To achieve the best of both worlds, we propose a generic modeling framework that can handle both direct and iterative forecasting. First we observe that as the time progresses, the dependency between the future weather state and the past weather state gradually decreases. As a result, we make the modeling assumption that for given $K_0$ and $\Delta k$, there exists a large enough $L$ such that $x^{(K_0+M\Delta k)} \perp\!\!\!\perp x^{(K_0)} \,|\, x^{(K_0+\Delta k)},\cdots,x^{(K_0+L\Delta k)}$ for all $M \geq L+1$. 

With this assumption, $P(x^{(K_0 + n\Delta k)}|y,[x^{(K_0 + j\Delta k)}]_{j<n}) \approx P(x^{(K_0 + n\Delta k)}|[x^{(K_0 + (n-j)\Delta k)}]_{j=1}^{L})$. Therefore, we can approximate Eq~\eqref{iterative_ar} with:
\begin{equation}
\small
    x^{(K_0 + N\Delta k)*} \approx \argmax_{x^{(K_0 + N \Delta k)}} \prod_{n=L}^N P(x^{(K_0 + n\Delta k)}|[x^{(K_0 + (n-j)\Delta k)}]_{j=1}^{L})
\end{equation}
where $x^{(K_0 + n\Delta k)}=f(y,x^{(K_0 + (n-L)\Delta k)},\cdots,x^{(K_0 + (n-1)\Delta k)})$ with direct model $f$ trained with the objective in Eq~\eqref{direct_obj}.

With the above approximation, we can achieve both direct and iterative forecasting with the same modeling framework.

\begin{figure*}[ht]
\begin{center}
\centerline{\includegraphics[width=\textwidth]{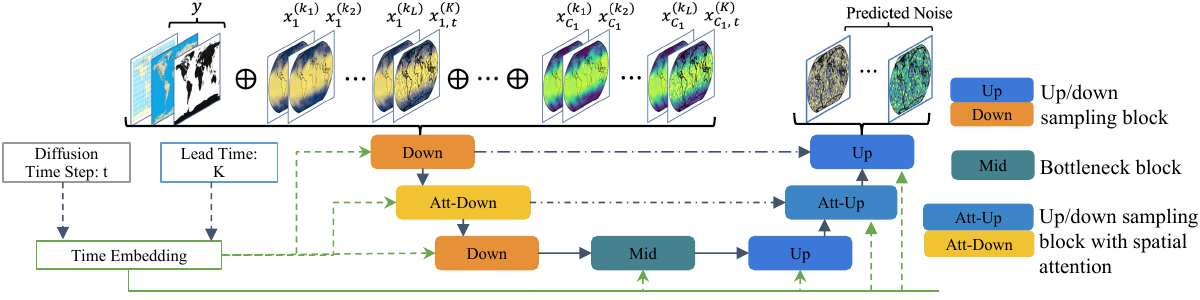}}
\caption{The overall diffusion model network architecture.
}
\label{denoiser_fig}
\end{center}
\vskip -0.2in
\end{figure*}

\subsection{Modeling with Conditional Diffusion}
To model this conditional distribution, we design a conditional diffusion model that explicitly incorporates the non-weather constants, the prior observations and the target lead time as input conditions along side with the noisy sample produced from the previous diffusion time step. The overall model architecture is shown in Figure~\ref{denoiser_fig} and the detail settings are in the \cref{denoise_network}.

More specifically, we denote the $c$-th channel in observed weather $x^{(k)}$ as $x^{(k)}_c$ and the $c$-th channel in the noisy sample at diffusion time step $t$ as $x^{(K)}_{c,t}$. We first perform a sequential channel-wise concatenation so that the same channels of all observed weather and the noisy sample as stacked together. Then we concatenate the non-weather constants on top of all the weather variables to form a vector 
$y\bigoplus (x^{(k_1)}_1,\cdots,x^{(k_L)}_1,x^{(K)}_{1,t})
\bigoplus \cdots \bigoplus (x^{(k_1)}_{C_1},\cdots,x^{(k_L)}_{C_1},x^{(K)}_{C_1,t})$ $\in \mathbb{R}^{H\times W\times (C_0+(L+1)C_1)}$. This vector together with the target lead time $K$ and the diffusion time step $t$ forms the inputs of our diffusion model.
To capture the periodicity of weather and to handle varying target lead time, we also create a Fourier embedding of the input lead time~\citep{transformer}.
Since weather patterns typically closely correlate with time, we explicitly treat the lead time variable as the ``class label'' of our conditional diffusion model.

While this predictability usually holds for predictions in the near future, it becomes less reliable as the forecast extends further into the future. Moreover, we can also observe that generally the future weather states are regarded as potential deviations from the previous observations and the posterior distribution of a weather state given a lead time is fairly concentrated around the prior weather distribution. As a result, our model can also gain benefits from learning the prior distribution of a weather state without conditioning on the lead time. To incorporate the benefit of learning from both the unconditional and conditional distribution,
we use classifier-free guidance~\citep{ho2022classifier} in our model training and sampling.
In the ablation study in \cref{cfg_weight}, we show that 
the larger the target lead time is, the smaller the guidance scale is required to produce a relatively accurate forecast.

Following the configurations described above, we train a diffusion model with total number of diffusion time steps $T$ with the following objective:
\begin{equation}
    L(\theta) = \mathbb{E}\left[\omega_i\|\epsilon_{\theta}(\sqrt{\alpha_{t}}x^{(K)} + \sqrt{1-\alpha_{t}}\epsilon; y, \Vec{x}_{ob}, t, K) - \epsilon\|^2 \right]
\end{equation}
where $\epsilon\sim N(0,I)$, $\epsilon_{\theta}$ is the diffusion model that predicts the noise for each time step, $\alpha_t$ is the predefined diffusion noise scale scheduling at diffusion time step $t$. Following~\citet{rasp2020weatherbench}, we also use $\omega_i$ as the latitude weighting factor for the $i$-th latitude to account for the projection scaling from the latitude-longitude grid to a 2-D Euclidean grid. For inference, we use DDIM~\citep{song2020denoising} sampling and details are provided in Section \ref{Experiments} and the appendix.

\subsection{Guided Weather Forecasting}
Besides directly sampling from the trained diffusion model, one can also easily incorporate various assumptions, constraints and guidance into the forecast generation.
using SDEdit, a technique developed by~\citet{meng2021sdedit}. The key idea of SDEdit is to inject a ``preliminary version of the target output'' in the middle of the diffusion process as guidance and continue removing the remaining noise to achieve an output that is both realistic and faithful to the input guidance. In the task of weather forecasting, the input guidance can be existing forecasting results that are easy to compute and follow certain physical constraints but are not necessarily accurate. SDEdit initializes the diffusion process from the neighborhood of the input guidance and gradually fills in the corrupted details of the forecast to be more precise based on the trained diffusion models. 
\begin{figure*}[ht]
\begin{center}
\centerline{\includegraphics[width=\textwidth]{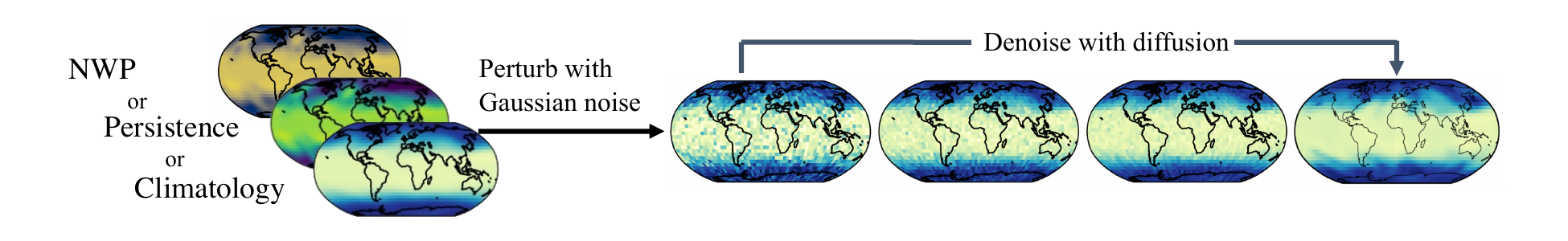}}
\caption{Generate weather forecasts from guidance of various types with our framework.}
\label{sdedit_illustration}
\end{center}
\vskip -0.2in
\end{figure*}

Mathematically speaking, let the existing forecast be $\hat{x}^{(K)}$, then using SDEdit at diffusion time step $t_0$ and $\hat{x}^{(K)}$ as input guidance is equivalent to imposing the assumption:
\[\sqrt{\alpha_{t_0}}x^{(K)} + \sqrt{1-\alpha_{t_0}}\epsilon \sim \mathcal{N}(\sqrt{\alpha_{t_0}}\hat{x}^{(K)}, (1-\alpha_{t_0})I)\]
with $\epsilon \sim N(0, I)$. 
With different choices of the starting diffusion time step $t_0$, users of our system can decide how much they would like to ``trust'' the diffusion model versus the existing forecast: Choosing a smaller $t_0$ suggests that the users believe that an accurate prediction lies in close proximity to the existing forecast $\hat{x}^{(K)}$, and therefore select a Gaussian distribution with smaller variance surrounding the scaled $\hat{x}^{(K)}$. Likewise, a larger $t_0$ implies that the users perceive the existing forecasting methodologies as suboptimal and therefore choose to explore more by selecting a larger variance and using the diffusion model to generate more content in the forecast. $t_0 = 0$ is equivalent to directly predicting with the existing forecast, and $t_0=T$ is equivalent to directly sampling from the diffusion model without any guidance. 

With different traditional weather forecasting methods, different $\hat{x}^{(K)}$ and the resulting outputs can provide unique properties. For example, this framework uses NWP predictions as a physical guide to ensure realistic outcomes; with persistence guidance, it manipulates how similar the next forecast is to the previous forecast. And with climatology guidance, it allows the generation of forecasts that approach the general physical pattern of weather over longer periods.

%% file: icml2024/tex/experiment.tex
\section{Experiments}\label{Experiments}

\subsection{Datasets}\label{datasets}
We trained the our model on a dataset built from the WeatherBench \citep{rasp2020weatherbench} which is the ERA5 reanalysis dataset \citep{hersbach2020era5} that was regridded to 5.625° ($H=32, W=64$) using a bilinear interpolation. We also subsampled the temporal resolution from 1 hour to 6 hours, corresponding to 00z, 06z, 12z and 18z times each day, where 00z means 00:00UTC in Zulu convention. To provide a basic demonstration of the task, we simply used $C_1=2$ two-dimensional input-output fields in the model: geopotential height at 500 hPa (Z500) and temperature at 850 hPa (T850). The Z500 fields are important in identifying the structure of midlatitude weather systems, and T850 is important in identifying frontal boundaries. We also input $C_0=3$ additional prescribed constant fields from Weatherbench: the land-sea mask, the topographic height, and the latitude. Our model is trained on the dataset from 1979 to 2015 for which the input is transformed by a min-max scaler and scaled to the range of $[-1, 1]$. Configurations are tuned based on the 2016 dataset and the results are validated on 2017 and 2018 datasets which are directly comparable to the baseline from WeatherBench.
\subsection{Implementation Details}
We choose $T=1000$ for all DMs. The $\Delta k$ is equal to 6 hours in all experiments. $C_{0}=3$ and $C_{1}=2$ for both DM iterative and DM direct as described in Section \ref{datasets}. Further details can be found in~\cref{experiment_details}.

In order to investigate the properties of different tasks, we train two separate lightweight model for iterative and direct tasks respectively even though the same framework can be applied to both tasks. For the remaining parts of this paper, we refer to the model trained for the iterative task as DM iterative and the one trained for the direct task as DM direct.
DM iterative consists of 4 consecutive observations $\Vec{x}_{\text{ob}} = \left[ x^{(0)}, x^{(1)}, x^{(2)}, x^{(3)}\right]$ with output lead time $K \in \{1, 2, 3, 4\}$. We choose to output a single lead time $K=1$ for the DM iterative model that is presented in \cref{baseline} and the following sections.
DM direct consists of 4 observations $\Vec{x}_{\text{ob}} = \left[ x^{(0)}, x^{(1)}, x^{(2)}, x^{(4)}\right]$
with output lead time $K \in \{1, 2, 4, 12, 20, 28, 36\}$.
\subsection{Evaluation Metrics}

We adopt two key metrics from WeatherBench~\citep{rasp2020weatherbench} to measure the quality of the forecasts: the mean latitude-weighted root mean squared error (RMSE) for assessing the error magnitudes, and the mean anomaly correlation coefficient (ACC) for assessing spatial anomaly pattern accuracy. RMSE measures the forecast precision, with lower values indicating greater accuracy. A perfect ACC score of 1 denotes absolute correlation with actual weather events, while 0 implies no predictive power. Unlike RMSE, ACC evaluates how well the forecast anomalies have represented the observed anomalies, which is crucial for determining if forecasts capture significant spatial weather variations or just approximate average weather conditions.

\subsection{Diffusion Forecast without Guidance}\label{baseline}
We first compare the performance of our proposed DM model to competing baselines without any input guidance.
\paragraph{Baselines}
We choose the same baselines mentioned in WeatherBench, which are briefly introduced as follows: (1) Persistence Forecast, which uses current conditions to predict the next day's weather; (2) Climatological Forecast, comprising Overall Mean Climatology (averaged over data from 1979 to 2016) and Weekly Mean Climatology, which is the averaging data for each of the 52 calendar weeks to account for seasonal variations; (3) Operational NWP Model, specifically the ECMWF's Integrated Forecast System, which requires substantial computational resources, exceeding 10,000 CPU cores; (4) Physical NWP Models, including the T42 Model with a resolution of 310 km, taking 270 seconds per forecast on 36 CPU cores, and the T63 Model with a resolution of 210 km, requiring 503 seconds per forecast on 36 CPU cores; (5) Linear Regression, which is used for both iterative 6-hour forecasts and direct 3-day and 5-day forecasts, trained on two weather states; and (6) a Convolutional Neural Network (CNN) featuring 5 layers, implemented for both iterative and direct forecasts, similar to the linear regression approach.

\begin{figure}[t]
\begin{center}
\centerline{\includegraphics[width=\columnwidth]{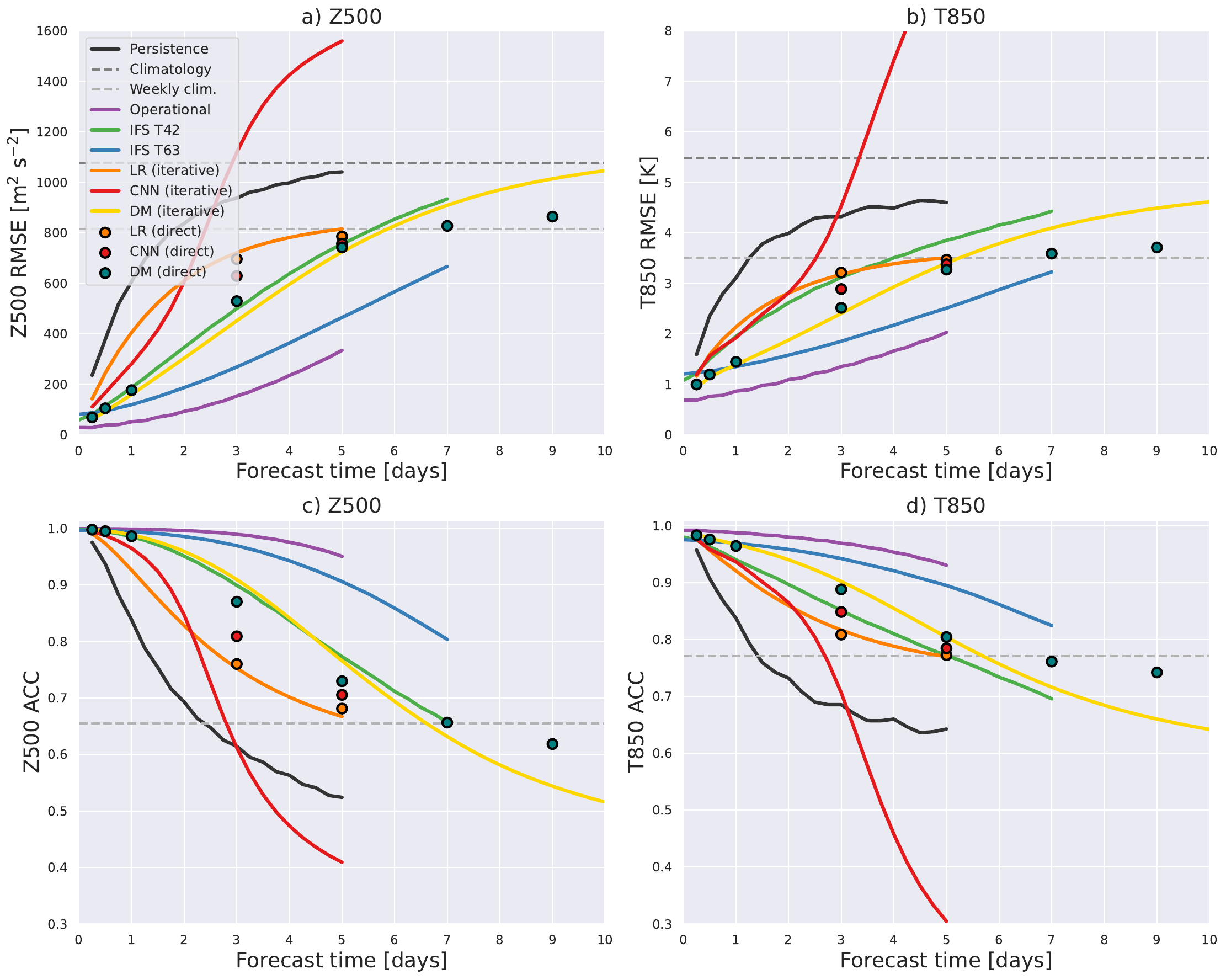}}
\caption{RMSE (a) and ACC (c) of 500 hPa geopotential and RMSE (b) and ACC (d) of 850 hPa temperature for different baselines at 5.625$^{\circ}$ resolution. Solid lines are iterative forecasts, while dots represent direct forecasts.}
\label{baseline_result}
\end{center}
\vskip -0.3in
\end{figure}

\paragraph{Results}
In \cref{baseline_result}a and \cref{baseline_result}b, both DM direct and DM iterative outperform persistence approaches and weekly climatology in terms of average global forecast RMSE in 500-hPa height and 850-hPa temperature fields, surpassing the latter for up to 7 days (DM direct) and 6 days (DM iterative). Our models also marginally outperform the T42 model for all forecast lengths, with the DM direct being slightly more accurate at 7 and 9 days. Similar trends are noted in T850's RMSE, with the DM models leading the T42 model by a day. Note that even though T63 and Operational models are superior to our DM models, our method still has advantages in terms of computational costs.

The ACC scores for Z500 (\cref{baseline_result}c) mostly mirror the RMSE rankings. DM models match the weekly climatology's ACC score at a forecast limit similar to the T42, but fall below weekly climatology's ACC just before 7 days. The DM iterative performs better than the DM direct in earlier forecasts. For T850 (\cref{baseline_result}d), our models have a clear advantage over the T42. Overall, our models' high ACC scores indicate that they effectively produce realistic spatial weather patterns with accurate disturbance amplitudes.

\subsection{Diffusion Guided by Realistic Forecast Processes}
Having compared the base performance of our DMs to the baselines, we can now explore how each guidance further improves the performance of the DM models. Specifically, we focus on three types of guidance: NWP guidance, persistence guidance and climatology guidance, and integrate them into the base iterative 6-hourly DM model. In general, combining DMs with different input guidance can provide various advatanges over the orignal models without any retraining.
\subsubsection{NWP Guidance}
Given an NWP forecast, we want to generate weather forecasts from the iterative DM that are faithful to the trusted NWP model but also retain the ability to draw samples from a potentially more accurate distribution. 
\begin{figure}[t]
\begin{center}
\centerline{\includegraphics[width=\columnwidth]{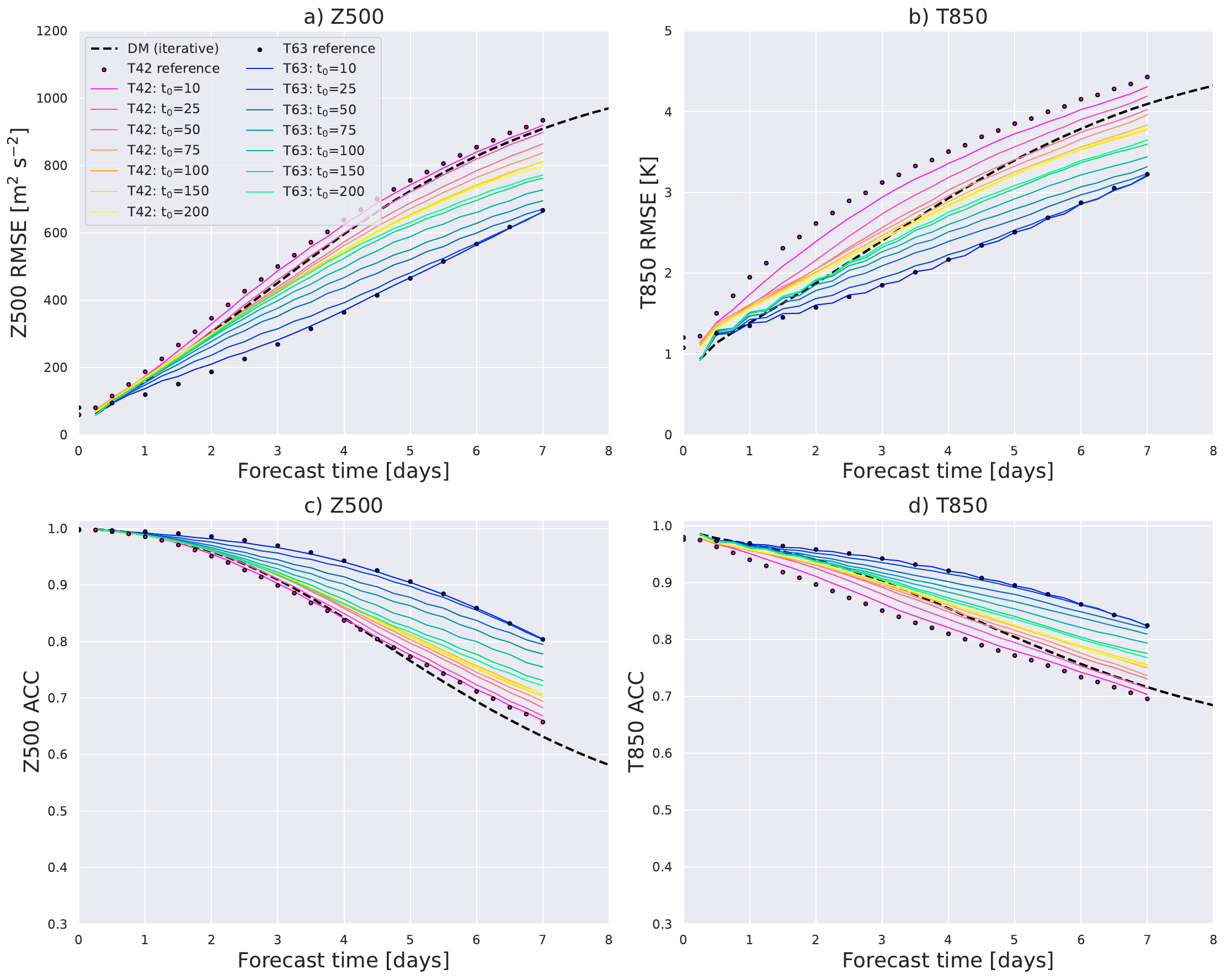}}
\caption{RMSE (a) and ACC (c) of 500 hPa geopotential and RMSE (b) and ACC (d) of 850 hPa temperature for incorporating IFS T42 and IFS T63 forecasts as guidance with different $t_0$
The dashed line is the basic 6-hourly DM iterative model, the magenta dot represents IFS T42 and the blue dot represents IFS T63.}
\label{NWP_guidance}
\end{center}
\vskip -0.4in
\end{figure}

\begin{figure*}[ht]
\begin{center}
\centerline{\includegraphics[width=\textwidth]{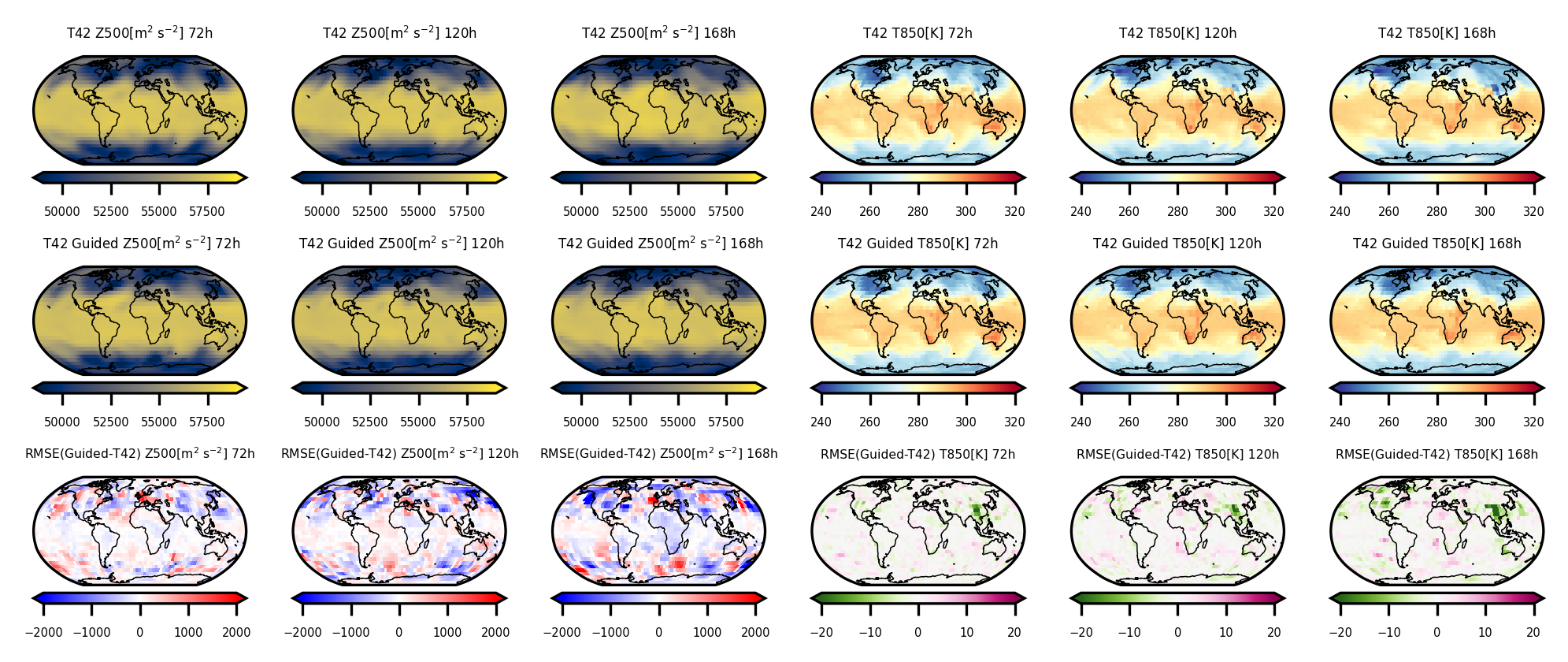}}
\caption{Example fields for 2017-12-24 12UTC initialization time. The top row shows the forecast produced by T42. The second row shows the forecast produced by the iterative DM after denoising from the T42 forecast perturbed by 200 steps of Gaussian noise. The last row show the difference of RMSE between the T42 guided iterative DM forecast and T42 forecast.}
\label{NWP_guidance_example}
\end{center}
\vskip -0.2in
\end{figure*}

We start with the T42 guidance (\cref{NWP_guidance}) which are lines with color gradients ranging from magenta to yellow. As shown in \cref{baseline}, both DM iterative and T42 have 6-hourly time steps with similar performance in both Z500 and T850. When we perturb the T42 guidance with relatively low noise (i.e., $t_0$ less than 25) which we trust more towards the output of the T42 model, the forecast produced by DM after denoising from the perturbed T42 forecast is similar to the performance with the T42 model. However, if we increase the perturbation of T42 output by up to 200 steps of Gaussian noise, we achieve overall better performance compared to both T42 and the basic DM iterative model, especially in longer lead times. The forecast skill for Z500 improved by over a day of lead time for day 7. As for T850, the performance of T42 is much worse compared to Z500, such that incorporating the T850 field generally penalizes the performance in an earlier lead time. But for longer lead time, with a perturbation of T42 output by 200 steps of Gaussian noise, the forecast skill for T850 also improved by over a day of lead time for day 7. An example forecast of T42, DM iterative with T42 guidance, and the difference of RMSE between the two predictions is shown in \cref{NWP_guidance_example}. 

As for T63 guidance with 12-hourly time steps which are lines with color gradients ranging from blue to cyan (\cref{NWP_guidance}), the DM iterative incorporates the T63 guidance once every two time steps. We find that perturbing the T63 guidance with stronger Gaussian noise overall penalizes the performance of the model relative to the forecast of T63. This is limited by the performance of the DM which is not robust enough to accurately predict the diffusion posterior mean or the noise component at any given step in the sampling process. Forecast samples with T63 guidance and further discussion about this limitations are in \cref{failure_analysis}. That being said, this framework is highly flexible as the user is allowed to adjust the amount of noise that represents their trust relative to the NWP model even for different time steps. Furthermore, it demonstrates the potential of blending in guidance with different time steps with the DM forecasts. 
\subsubsection{Persistence Guidance}\label{section_persist}
We use the DM iterative output for the previous lead time as te persistence guidance and aim to generate forecasts that resemble the previous outputs. The key idea is that, with this prior imposed, it is less likely to obtain a diffusion posterior mean with large deviations during the sampling process.    
\begin{figure}[ht]
\begin{center}
\centerline{\includegraphics[width=\columnwidth]{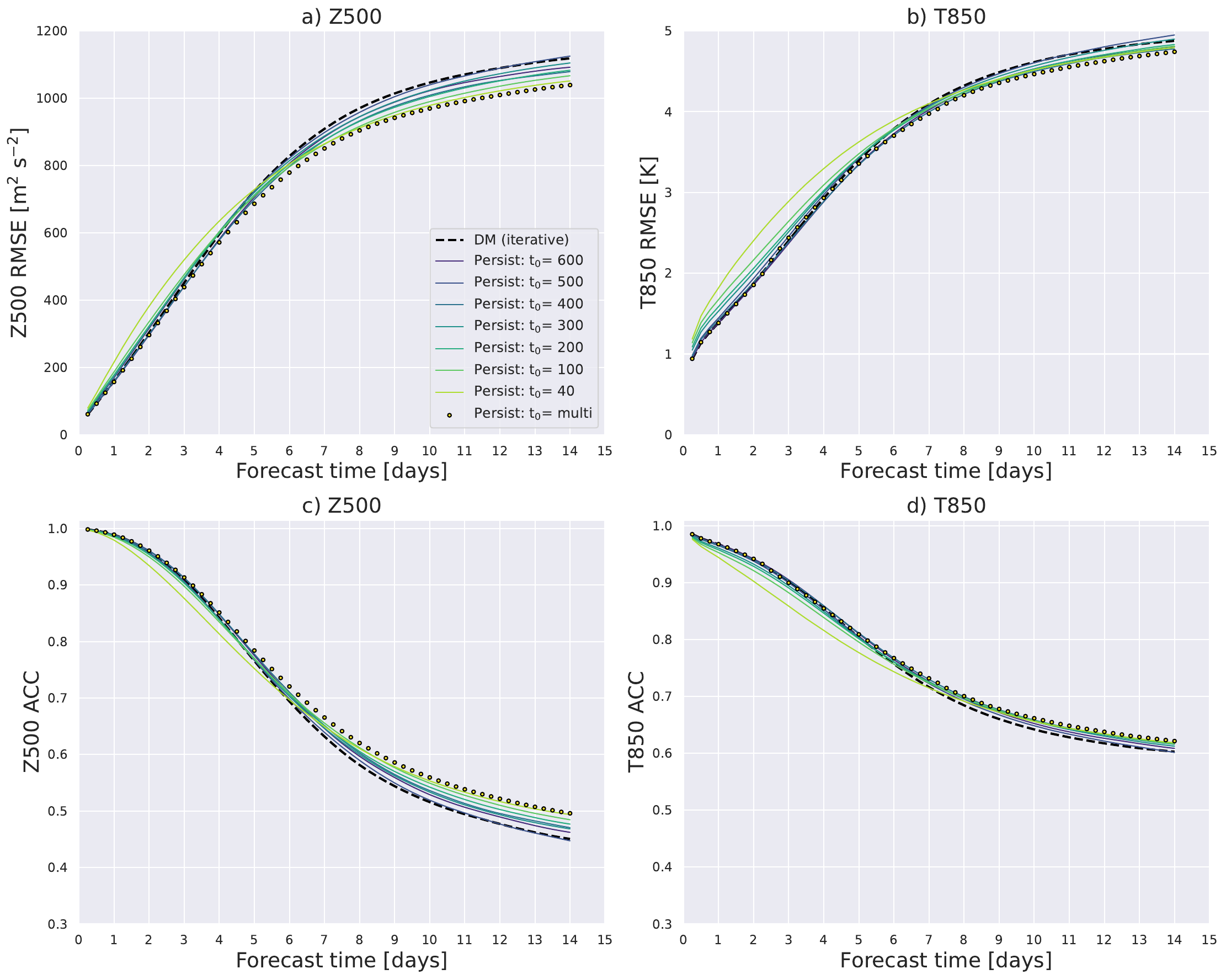}}
\caption{RMSE (a) and ACC (c) of 500 hPa geopotential and RMSE (b) and ACC (d) of 850 hPa temperature for incorporating previous forecasts from the DM iterative model itself as guidance 
with different $t_0$.
The dashed line is the basic 6-hourly DM iterative model, and the yellow dot represents an optimized combination for the amount of perturbation for each lead time. The configuration of ``Persist $t_0=$ multi'' is in \cref{persist_climo_t0_config}}
\label{Persistence_guidance}
\end{center}
\vskip -0.3in
\end{figure}

The persistence guidance generally improves the forecast skill in longer lead times. Looking at the RMSE and ACC of Z500 (\cref{Persistence_guidance} (a) and (c)), DM iterative performs relatively well for up to 4 days when compared to other persistence guidance. Nevertheless, incorporating the previous forecasts and perturbing with relatively strong noise (i.e. 500 and 600 steps) in early lead times still slightly improves the skill of the forecast. Reducing the perturbation added to the previous forecast, which lets the DM assume the forecast varies less than its previous forecast, in a longer lead time beyond 5 days, reduces the model error growth. These characteristics are also found in the T850 field (\cref{Persistence_guidance}b,d). However, because T850 has a larger spatial variation in time, the performance advantages that persistence guidance brings are not as significant as in the Z500 field, such that showing more trust toward previous forecast output as guidance more heavily penalizes performance in earlier lead times compared to the Z500 field. Finally, note that the yellow dot in \cref{Persistence_guidance} is a combination of the perturbation added to the previous forecast with smaller perturbations in shorter lead times and gradually larger perturbations in longer lead times, which boosts the forecast performance at all lead times compared to the basic DM iterative model. Forecast samples and configurations are presented in \cref{persist_climo_examples}

\subsubsection{Climatology Guidance}
For the climatology guidance task, we want to generate weather forecasts from the iterative DM model that resemble the climatology, which is the weather conditions averaged over a period of time.
Imposing climatology guidance reduces the likelihood of obtaining a biased weather prediction that deviates from the general climate states such as seasonality, which many DL models struggle with in free-running experiments.

\begin{figure}[ht]
\begin{center}
\centerline{\includegraphics[width=\columnwidth]{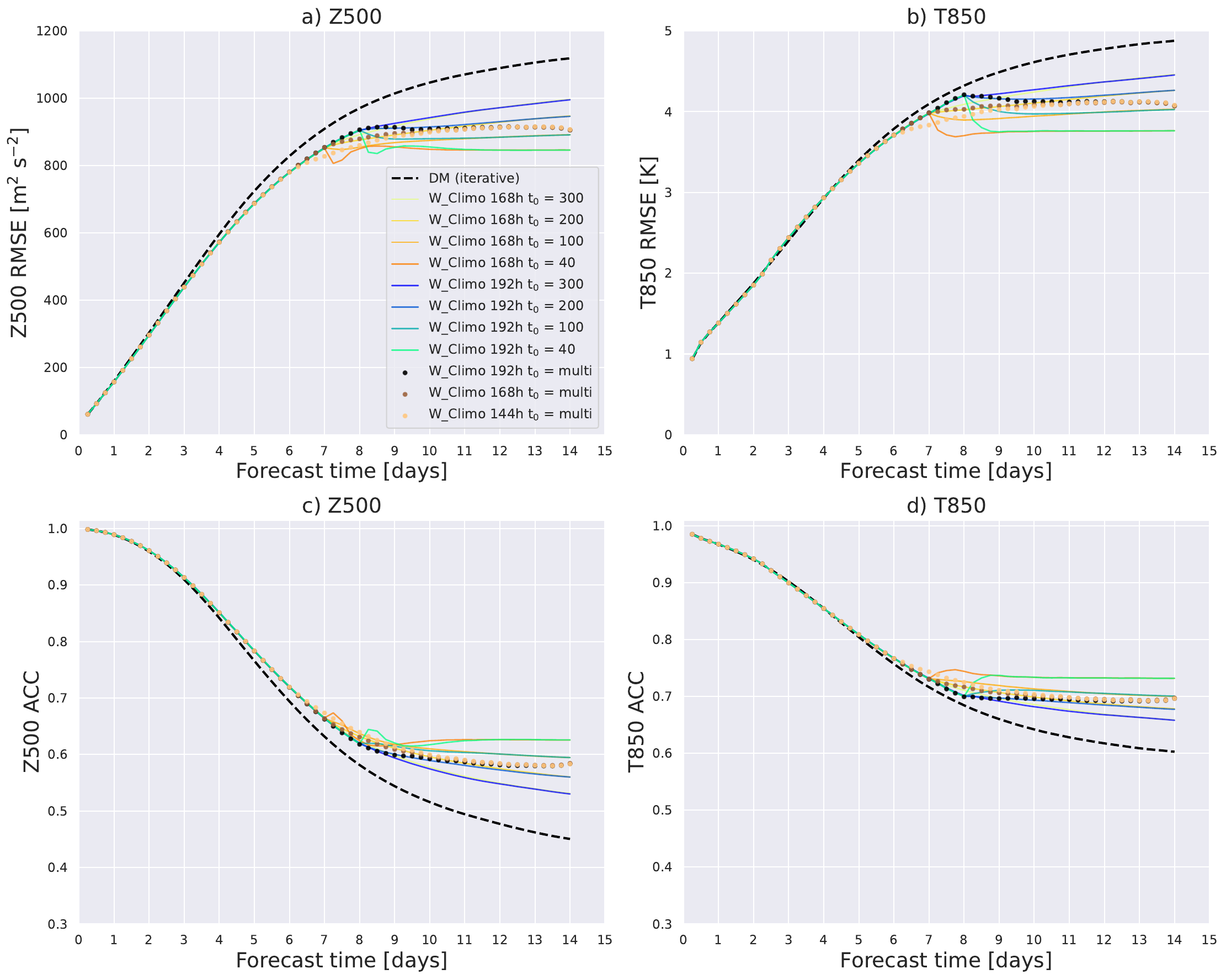}}
\caption{RMSE (a) and ACC (c) of 500 hPa geopotential and RMSE (b) and ACC (d) of 850 hPa temperature for incorporating ``Persist: multi'' from \cref{Persistence_guidance} and weekly climatology (W\_Climo) to the DM iterative model as guidance
with different $t_0$.
The dashed line is the basic 6-hourly DM iterative model, and the ``multi'' in the legend denotes incorporating different $t_0$ for each lead time with the configuration of ``W\_Climo 144h $t_0 = $ multi'' shown in \cref{persist_climo_t0_config}}
\label{Climo_guidance}
\end{center}
\vskip -0.3in
\end{figure}
We use the weekly climatology as it captures seasonal variation compared to overall mean climatology. The weekly climatology generally improves forecasts of the model near or beyond the lead time of the weekly climatology skill which such skill was shown in baseline (\cref{baseline}). Before starting to apply the weekly climatology guidance (\cref{Climo_guidance}), persistence guidance is applied to all runs based on the optimized combination for the amount of perturbation for each lead time in \cref{section_persist}. 

\cref{Climo_guidance} (a) and (c) show the performance of the Z500 field, with examples that applied weekly climatology guidance.
Incorporating weekly climatology guidance starting from day 8 with $t_0=300$ can already achieve an RMSE $<1000\si{m^{2}.s^{-2}}$ in Z500 at day 14 with an ACC of greater than 0.5. Similar performance gains can also be found for the T850 field. For experiments that impose stronger weekly climatology guidance starting either from day 7 or day 8, a more abrupt adjustment to the results can be observed in the figure. To account for this abruption, we gradually decrease $t_0$ as the lead time increases, and label those runs as ``$t_0 = $ multi''. An example forecast is presented for ``W\_Climo 144h $t_0 = $ multi'' in \cref{climo_examples}. Note that incorporating strong weekly climatology guidance can lead to the oversight of important diurnal variations in weather forecasts, thereby diminishing their overall relevance and accuracy. That being said, users also have the flexibility to incorporate other types of smoothing climatology with different kernel sizes to suit their needs.     

\subsection{Ablation Study}

In our experiments, we also find that the DM's performance is sensitive to the number of DDIM sampling steps and classifier-free guidance weight, which is rarely documented in the related literature. Our findings indicate that for the specific model and task of predicting two weather states based on two observed states, as few as 12 steps of sampling are sufficient to obtain optimal performance. Additionally, our results suggest that a decrease in the classifier-free guidance weight is necessary to maintain relatively good performance. The detailed analysis about these findings are in \cref{sensitivity_sampling_steps} and \cref{cfg_weight}. Lastly, \cref{failure_analysis} offers a brief analysis of the challenges encountered when incorporating the T63 guidance and discusses the limitations of the SDEdit framework in weather forecasting applications.

%% file: icml2024/tex/conclusion.tex
\section{Conclusion}
We introduce a novel unified framework that uses conditional diffusion models for weather forecasting. Our method can generate predictions across various future time points with great flexibility for both iterative and direct forecasts. In addition, we enhance our method by integrating SDEdit, which is adaptable to suit diverse weather forecasting needs by providing different physical guidance without retraining the models. These guidance include numerical weather predictions, persistence and climatology, which are methods that generally are used in the weather forecasting community. This framework allows users achieve trustworthy and competitive results in a low-resource setting and is adjustable based on their needs and preferences. In addition, the persistence and climatology guidance also improve the long-term forecast stability of our diffusion models. Our framework holds the potential to integrate a broader spectrum of guidance methods, paving the way for more trustworthy forecasting solutions with diffusion models.

%% file: icml2024/tex/impact_statement.tex
\section{Impact Statement}
Our work proposes a unified framework using diffusion models for future weather forecasting across various periods. We further explore the potential of combining realistic forecast processes such as the traditional numerical weather forecast (NWP) to guide the weather forecasts generated by the diffusion models. Specifically, our approach allows users to balance their trust in NWP and diffusion models in producing forecasts. On the bright side, incorporating NWP guidance into diffusion model forecasts enables users to anticipate weather system evolution based on prior experiences with such guidance. This facilitates more applications of DL-based models and improves the trustworthiness of these models in weather forecasting. However, it is crucial to remain cautious about the risks associated with generative AI in weather forecasting. Despite our method offering controllable forecast generation to minimize unrealistic weather predictions, there is still a possibility of generating plausible but incorrect forecasts, a challenge that even NWP models have been striving to overcome for decades. Continuous monitoring and cross-validation with other forecast sources are essential to ensure the reliability of our method.

%% file: icml2024/tex/acknowledge.tex
\section{Acknowledgement}
We would like to acknowledge high-performance computing support from Casper \cite{cisl2019cheyenne} and Derecho: HPE Cray EX System  (\url{https://doi.org/10.5065/qx9a-pg09}) provided by NCAR's Computational and Information Systems Laboratory, sponsored by the National Science Foundation.

%% file: icml2024/tex/appendix.tex
\onecolumn
\section{Experiment Details}\label{experiment_details}
\subsection{Diffusion Model Architecture}
\label{denoise_network}

Our diffusion model network architecture (\cref{denoiser_fig}) follows the settings in \citealt{ho2020denoising}, featuring channel depths of 64, 128, and 256 at each respective level. To condition on the noise level $t$ and lead time $K$, we follow the implementation in \cite{ho2020denoising} which creates sinusoidal timestep embeddings. The noise level $t$ is scaled within the range [0,1] whereas the lead time $K$ label adds a positive constant to scale greater than a minimum of 1. The sinusoidal timestep embeddings of $t$ and $K$ are passed through a 2-layer MLP and summed together. Each of the ResNet blocks in the block applies a further MLP layer to the summation of the aforementioned timestep embeddings which are then added to the hidden state of the ResNet block. 

\subsection{Diffusion Hyperparameter Settings}
The total number of diffusion steps is identical to DDPM~\citep{ho2020denoising} in which $T = 1000$. Similarly, the $\beta_{t}$ starts at 0.0001 and ends at 0.02 with the linear scheduling. We train our model using classifier-free guidance with $35\%$ conditional training and $65\%$ unconditional training.

\subsection{Further Implementation Details}
\paragraph{Training} Both DM direct and DM iterative are trained on a single Nvidia A100 with a batch size of 32. Each model training takes approximately 20 hours until no noticeable performance improvement based on the RMSE of the sampled forecast from the validation dataset.
\paragraph{Inference} The DM iterative with $K=1$, which is shown in the experiments, requires 56 iterative sampling steps to achieve a 14-day (336h) forecast. We select 12 as the number of DDIM sampling steps and 1.75 as the classifier-free guidance weight based on ablation study in \cref{sensitivity_sampling_steps} and \cref{cfg_weight}. With these settings, generating a 14-day forecast is accomplished in 1.1 seconds on a single Nvidia A100. The advantage in speed of using DM direct is much more significant here since this model can produce a 9-day forecast in a single run with 4 sampling steps. As for SDEdit with $t_0=200$, we used 40 DDIM sampling steps to generate each guided forecast.

\section{Ablation Study}
\subsection{The Effect of The Numbers of Sampling Steps}\label{sensitivity_sampling_steps}

\begin{figure}[t]
\begin{center}
\centerline{\includegraphics[width=0.8\columnwidth]{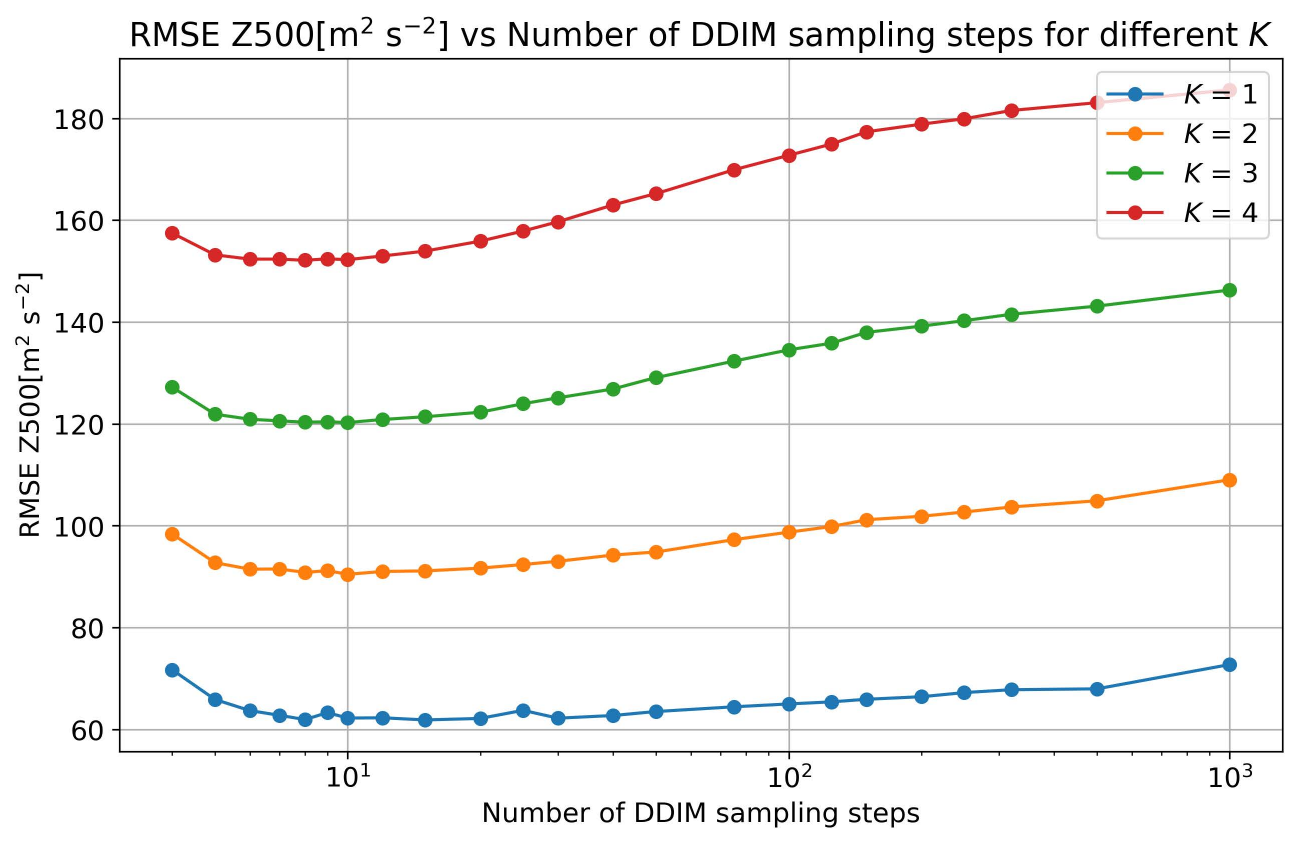}}
\caption{The effect of different number of DDIM sampling steps on RMSE Z500 for DM iterative with different lead times $K$. The classifier-free guidance weight is fixed as 1 in this experiment.}
\label{iter_sample_sense}
\end{center}
\end{figure}

\begin{figure}[t]
\begin{center}
\centerline{\includegraphics[width=0.8\columnwidth]{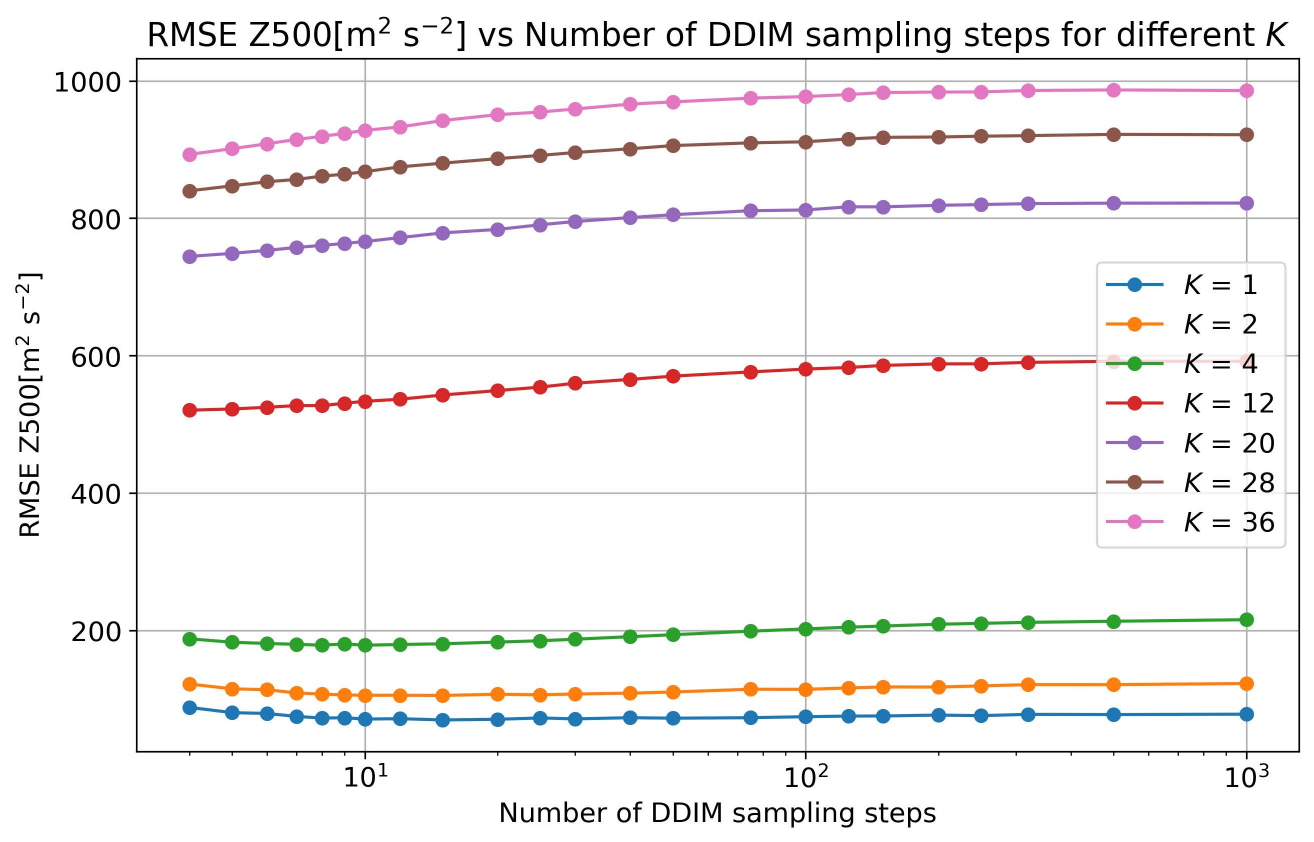}}
\caption{The effect of different number of DDIM sampling steps on RMSE Z500 for DM direct with different lead times $K$. The classifier-free guidance weight is fixed as 1 in this experiment.}
\label{direct_sample_sense}
\end{center}
\end{figure}

Our lightweight diffusion model, designed to forecast two weather variables using two input weather variables, demonstrates that the RMSE of Z500 reaches its minimum within 12 DDIM sampling steps. This finding holds true for both the DM iterative (\cref{iter_sample_sense}) and DM direct (\cref{direct_sample_sense}) methods, which differ in prior weather conditions and lead times. Notably, when the lead time $K$ is less than 4, the RMSE of Z500 is minimized at approximately 10 steps. For the DM direct method, a greater value of K results in a need for fewer DDIM sampling steps to minimize the RMSE, with 4 steps being optimal in our current setup when $K$ is greater than 12.

Typically, it is challenging to achieve high-quality results from high-resolution natural images with a limited number of sampling steps, due to the high dimensionality of the data distribution. However, in our experiment, the model we employ is specifically trained to estimate noise for sypnotic scale variables like Z500 and T850, which exhibit less spatial variation also with a relatively low resolution of \(32 \times 64\) latitude-longitude grids. This suggests that the dimensionality of our data is considerably lower than that of more complex natural images, and therefore it requires fewer DDIM steps to achieve high quality predictions. A recent study by DeepMind~\citep{price2023gencast} also demonstrates that their model can accurately sample weather conditions across more than 80 variables at a spatial resolution of \(180 \times 360\) latitude-longitude grids with 20 steps using the DPM Solver++ 2S \cite{lu2022dpm}, which supports our findings.

\subsection{The Effect of The Classifier-Free Guidance Weights}\label{cfg_weight}

\paragraph{DM Iterative}
\begin{figure}[ht]
\begin{center}
\centerline{\includegraphics[width=0.9\columnwidth]{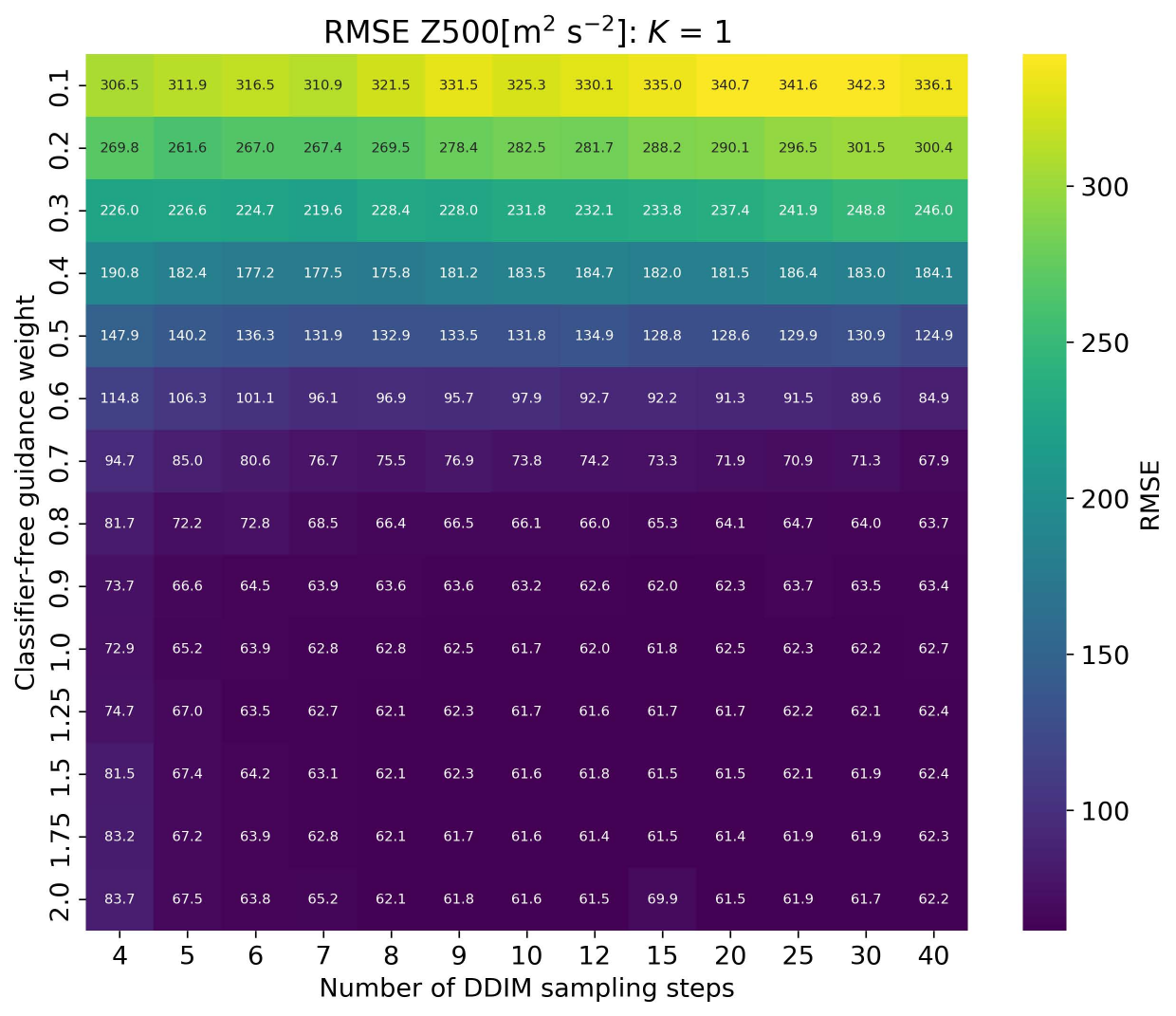}}
\caption{The effect of different classifier-free guidance weights and different number of DDIM sampling steps on RMSE Z500 for DM iterative with $K=1$.}
\label{iter_cfg_k=1}
\end{center}
\end{figure}

\begin{figure}[ht]
\begin{center}
\centerline{\includegraphics[width=0.9\columnwidth]{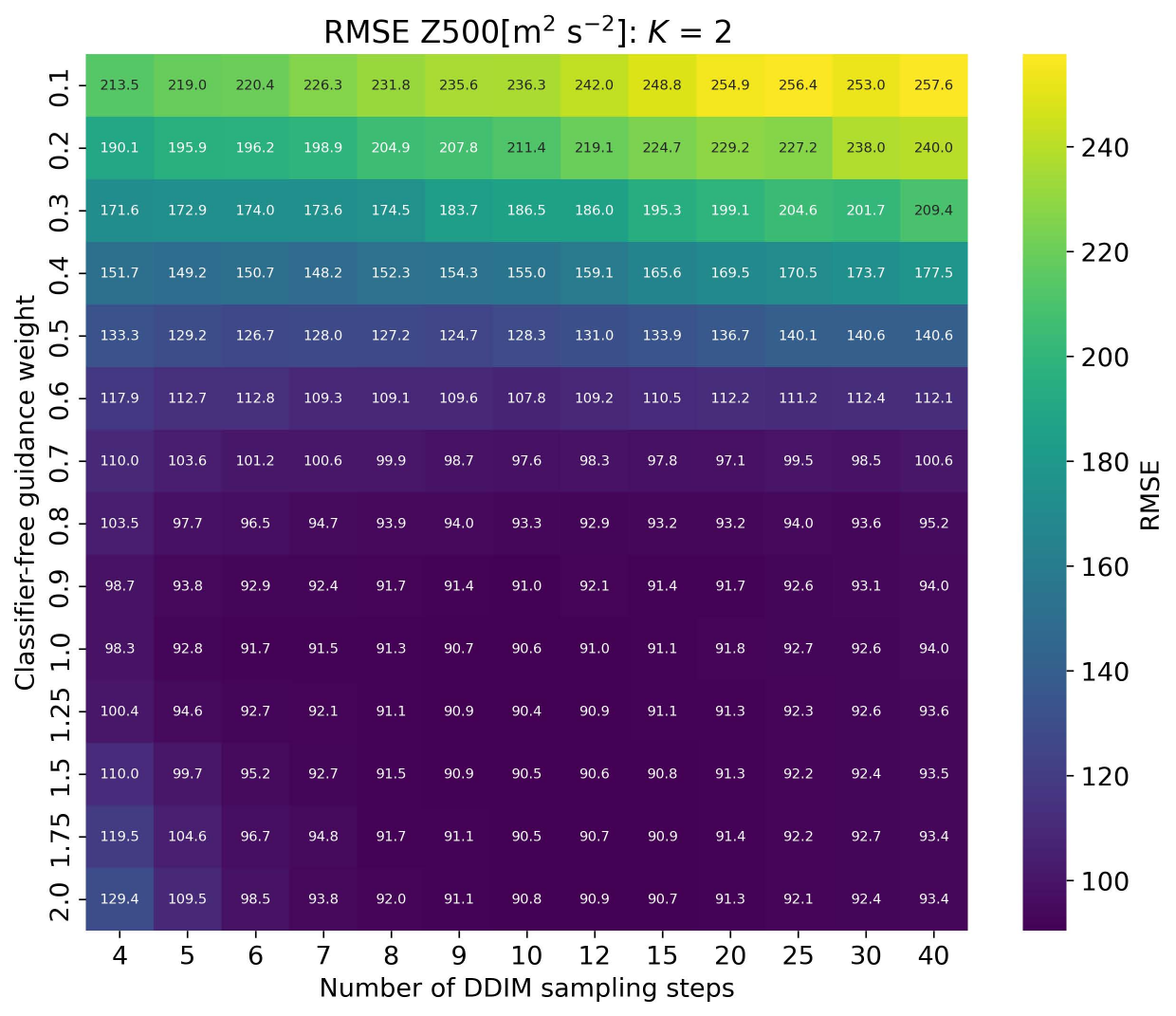}}
\caption{The effect of different classifier-free guidance weights and different number of DDIM sampling steps on RMSE Z500 for DM iterative with $K=2$.}
\label{iter_cfg_k=2}
\end{center}
\end{figure}

\begin{figure}[ht]
\begin{center}
\centerline{\includegraphics[width=0.9\columnwidth]{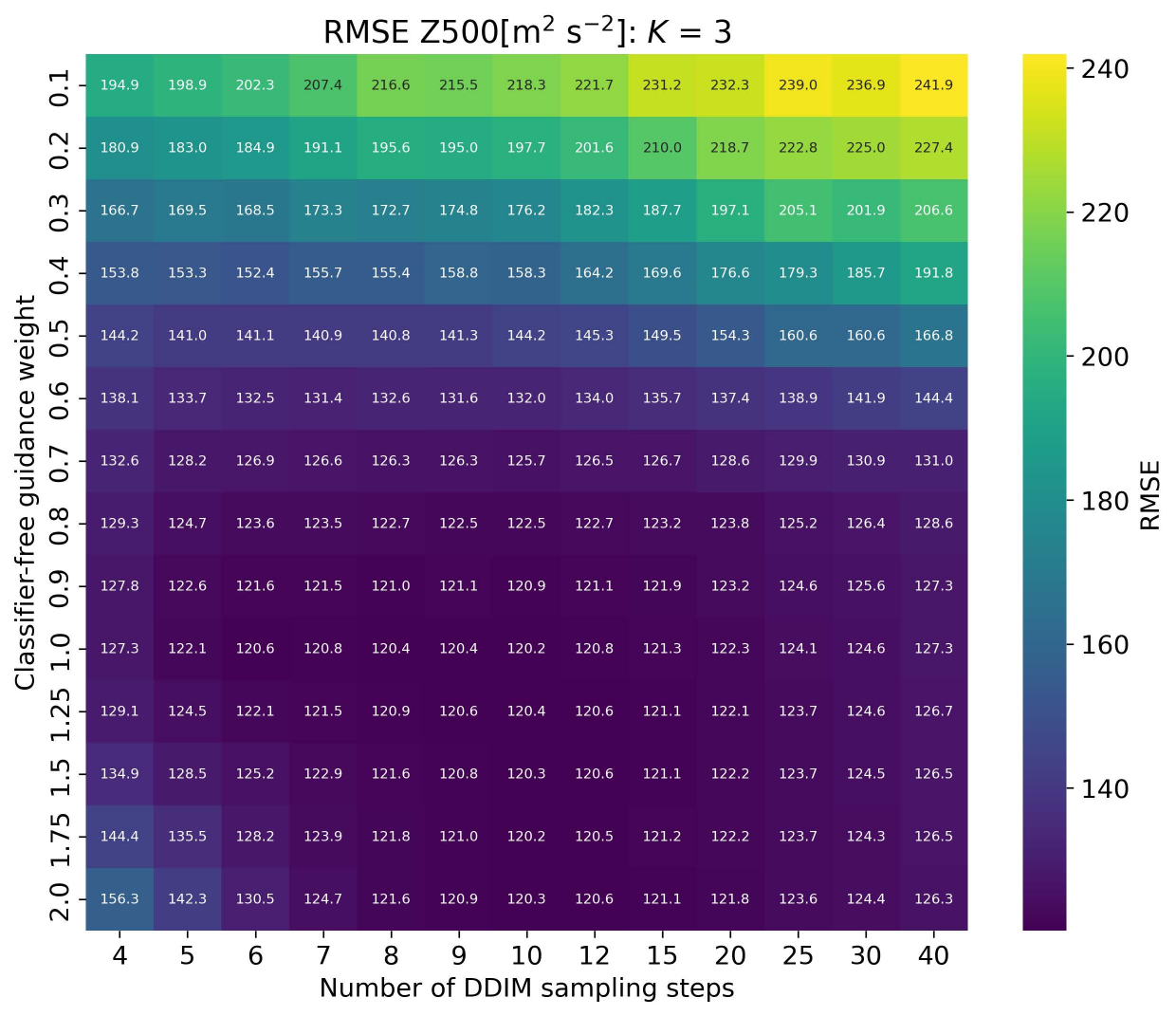}}
\caption{The effect of different classifier-free guidance weights and different number of DDIM sampling steps on RMSE Z500 for DM iterative with $K=3$.}
\label{iter_cfg_k=3}
\end{center}
\end{figure}

\begin{figure}[ht]
\begin{center}
\centerline{\includegraphics[width=0.9\columnwidth]{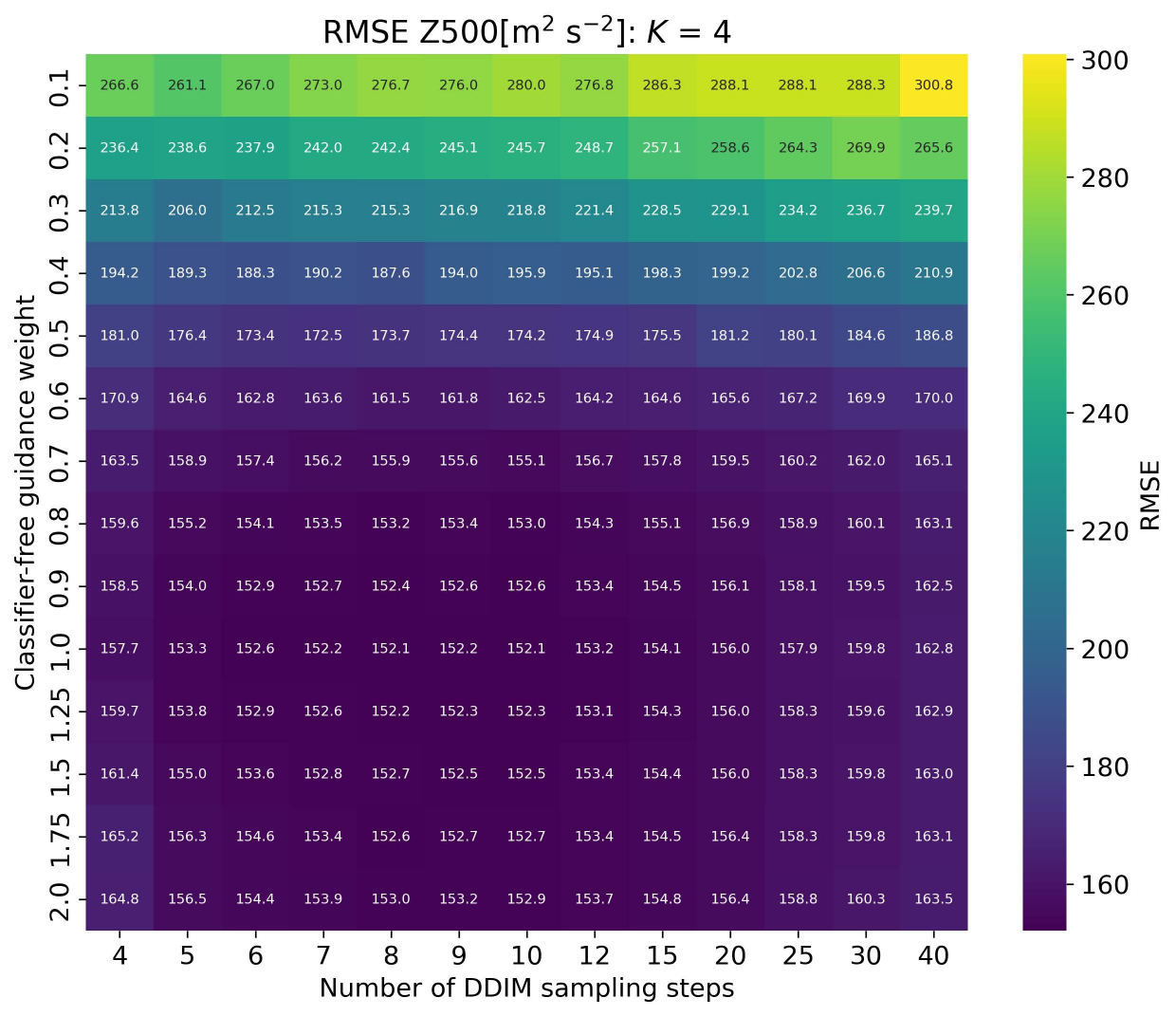}}
\caption{The effect of different classifier-free guidance weights and different number of DDIM sampling steps on RMSE Z500 for DM iterative with $K=4$.}
\label{iter_cfg_k=4}
\end{center}
\end{figure}

Our analysis demonstrates the impact of varying the classifier-free guidance weight on the Z500 RMSE for the DM iterative model across different numbers of DDIM sampling steps, as illustrated in Figures \ref{iter_cfg_k=1} to \ref{iter_cfg_k=4}. We observe that for the DM iterative model, a classifier-free guidance weight above 1 and below 2 consistently yields a lower RMSE. Furthermore, it appears that as lead time $K$ increases, a lower classifier-free guidance weight is needed to attain a similarly low RMSE. This trend shows the necessity of choosing the classifier-free guidance weight with respect to the lead time $K$ to optimize the model's performance.
 
\paragraph{DM Direct}

\begin{figure}[ht]
\begin{center}
\centerline{\includegraphics[width=0.9\columnwidth]{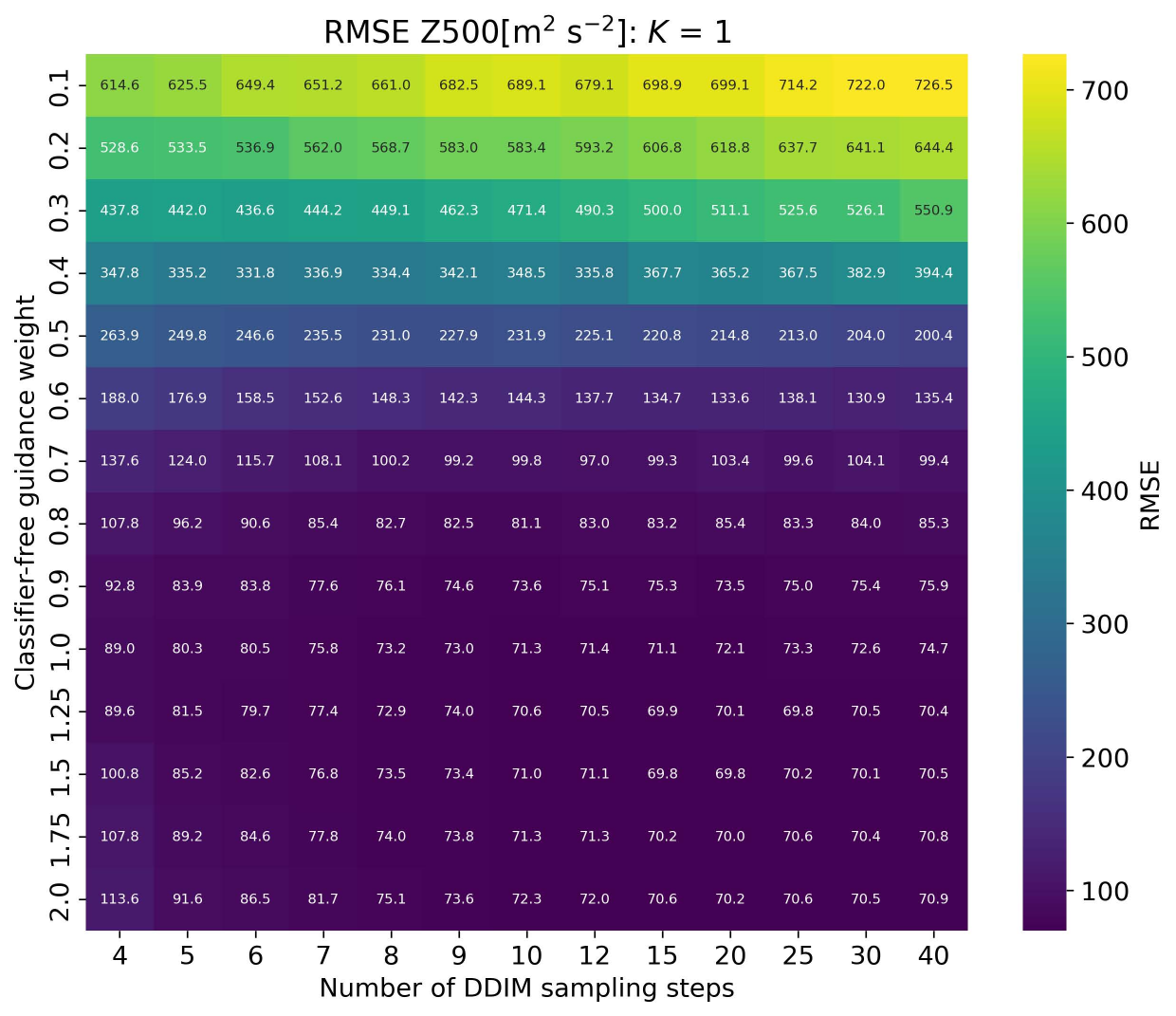}}
\caption{The effect of different classifier-free guidance weights and different number of DDIM sampling steps on RMSE Z500 for DM direct with $K=1$.}
\label{direct_cfg_k=1}
\end{center}
\end{figure}

\begin{figure}[ht]
\begin{center}
\centerline{\includegraphics[width=0.9\columnwidth]{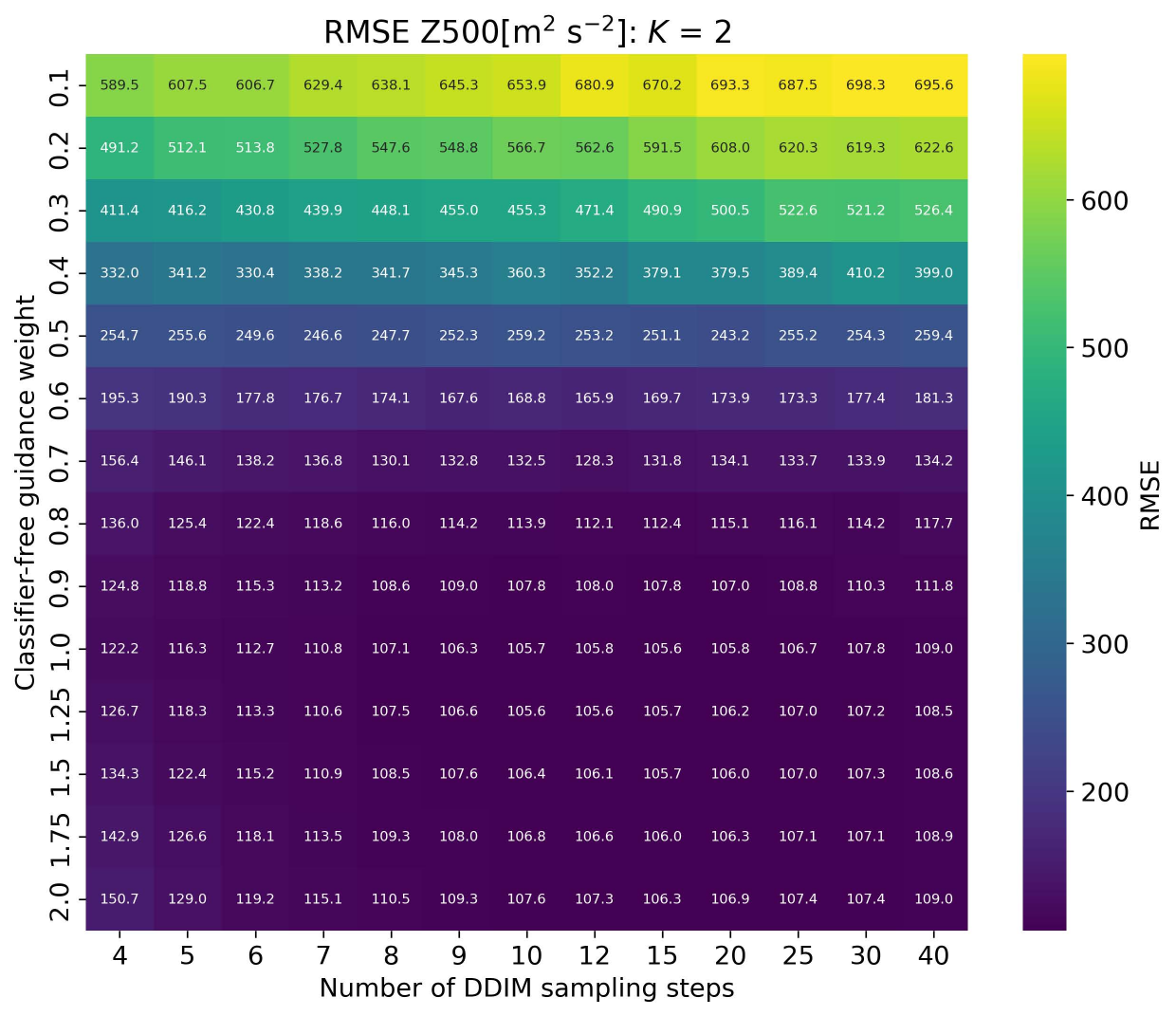}}
\caption{The effect of different classifier-free guidance weights and different number of DDIM sampling steps on RMSE Z500 for DM direct with $K=2$.}
\label{direct_cfg_k=2}
\end{center}
\end{figure}

\begin{figure}[ht]
\begin{center}
\centerline{\includegraphics[width=0.9\columnwidth]{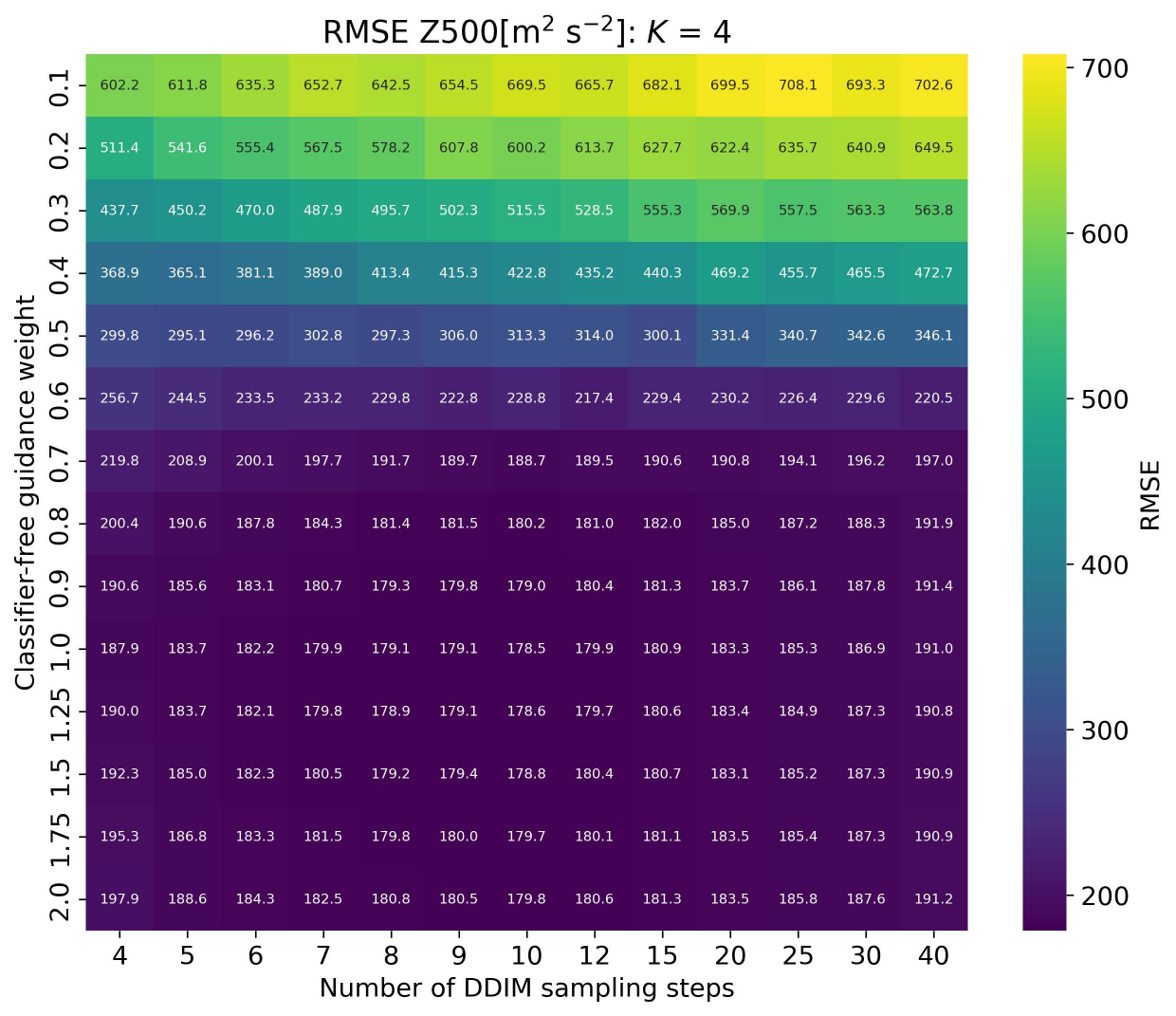}}
\caption{The effect of different classifier-free guidance weights and different number of DDIM sampling steps on RMSE Z500 for DM direct with $K=4$.}
\label{direct_cfg_k=4}
\end{center}
\end{figure}

\begin{figure}[ht]
\begin{center}
\centerline{\includegraphics[width=0.9\columnwidth]{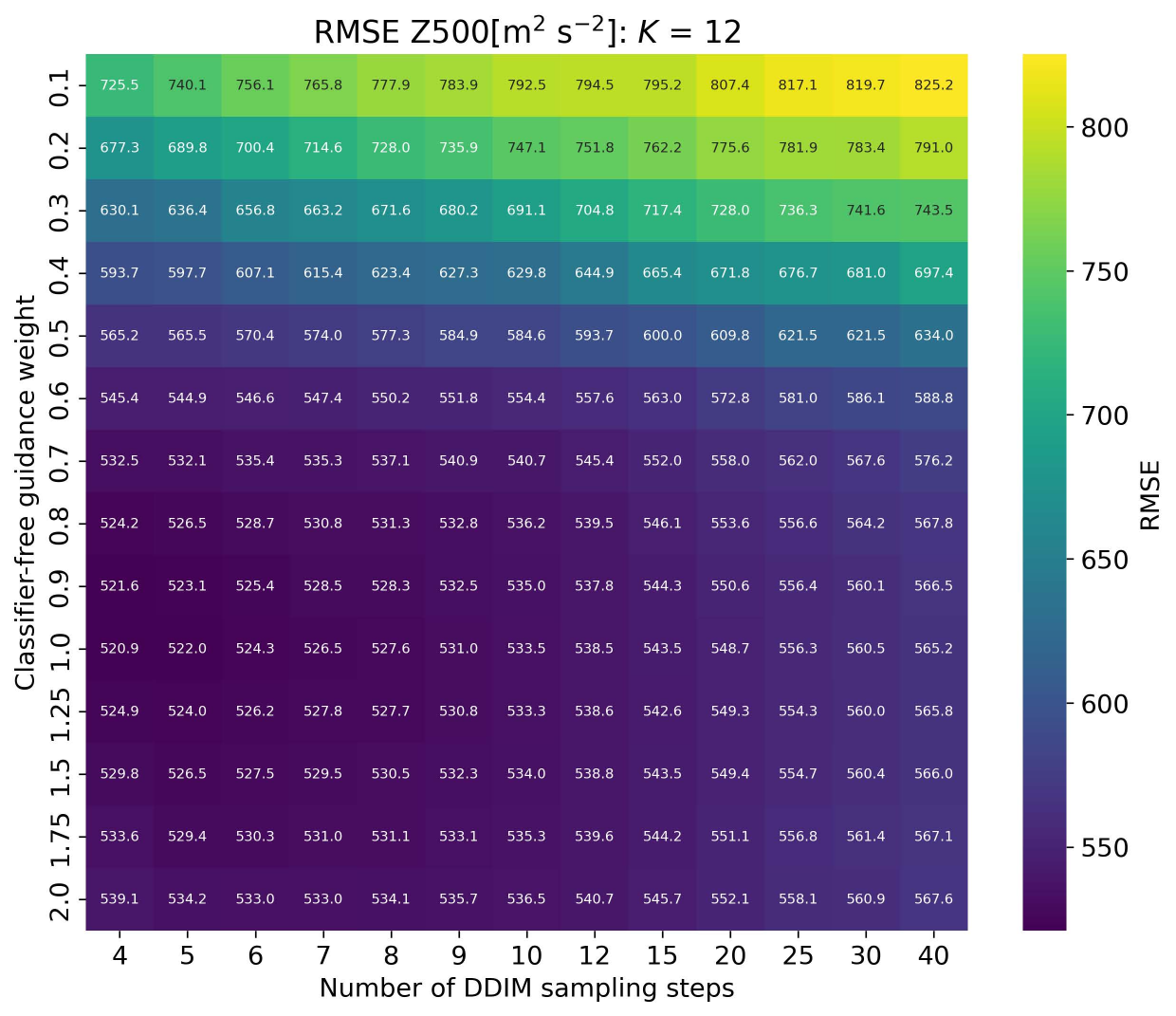}}
\caption{The effect of different classifier-free guidance weights and different number of DDIM sampling steps on RMSE Z500 for DM direct with $K=12$.}
\label{direct_cfg_k=12}
\end{center}
\end{figure}

\begin{figure}[ht]
\begin{center}
\centerline{\includegraphics[width=0.9\columnwidth]{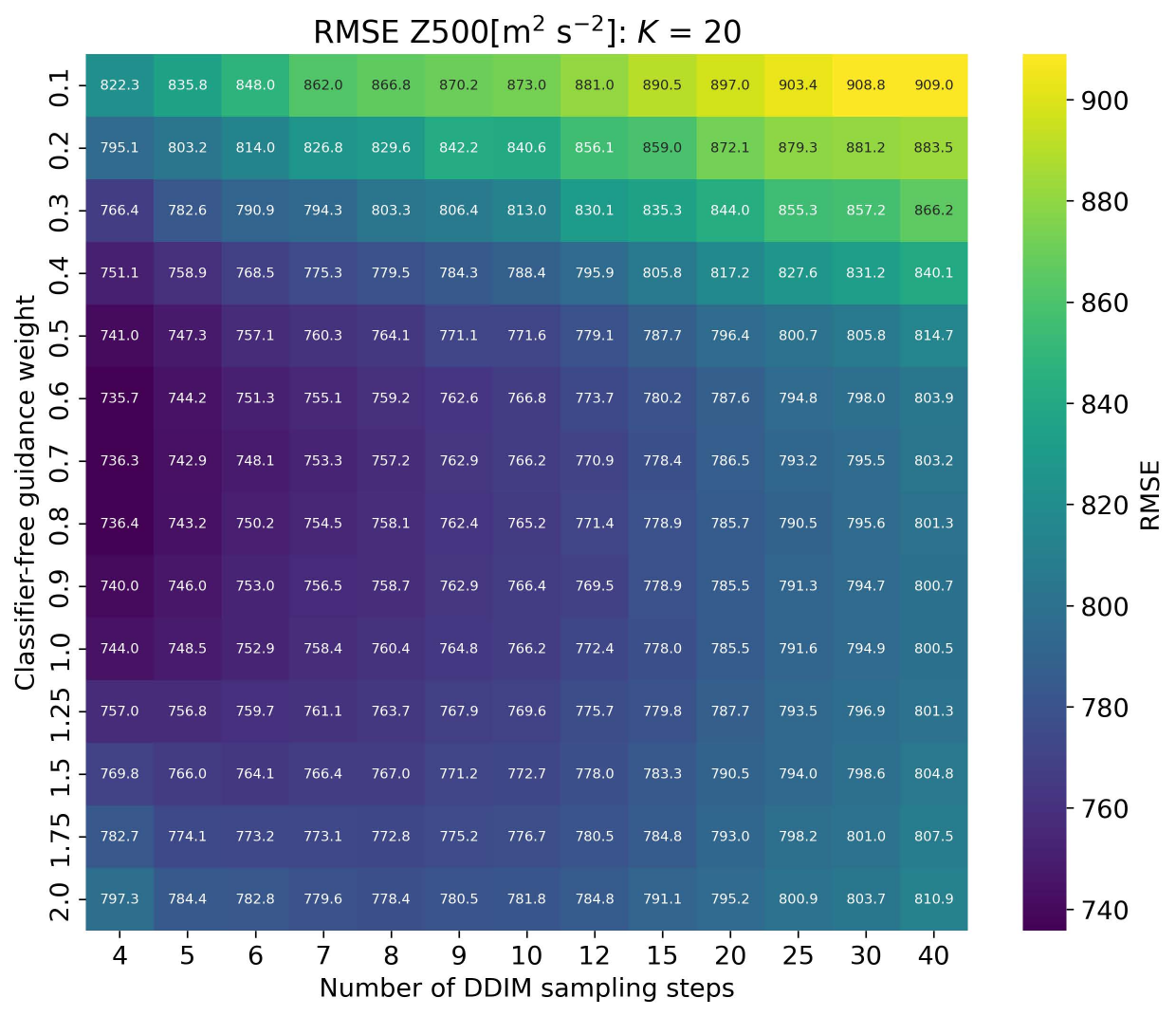}}
\caption{The effect of different classifier-free guidance weights and different number of DDIM sampling steps on RMSE Z500 for DM direct with $K=20$.}
\label{direct_cfg_k=20}
\end{center}
\end{figure}

\begin{figure}[ht]
\begin{center}
\centerline{\includegraphics[width=0.9\columnwidth]{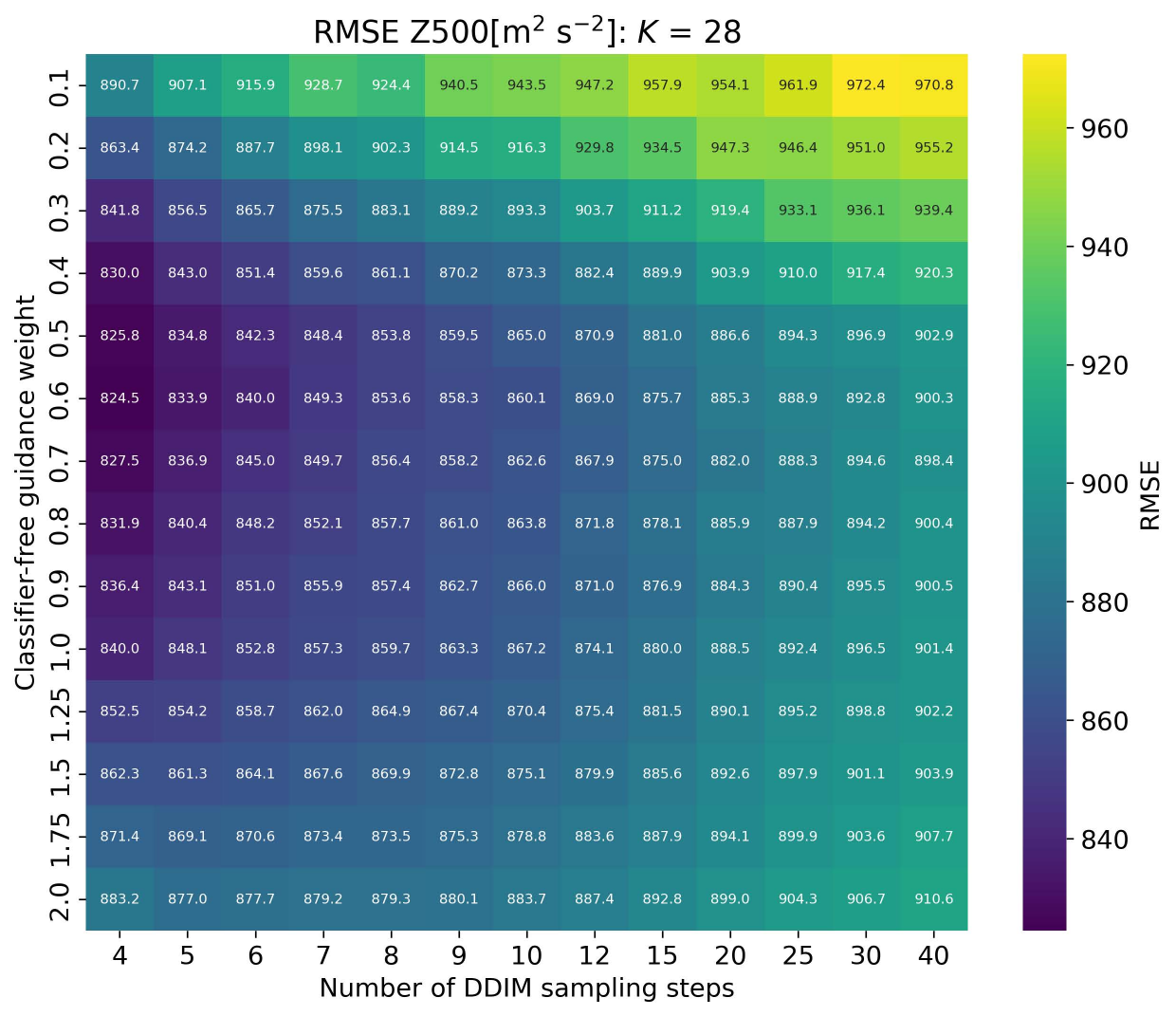}}
\caption{The effect of different classifier-free guidance weights and different number of DDIM sampling steps on RMSE Z500 for DM direct with $K=28$.}
\label{direct_cfg_k=28}
\end{center}
\end{figure}

\begin{figure}[ht]
\begin{center}
\centerline{\includegraphics[width=0.9\columnwidth]{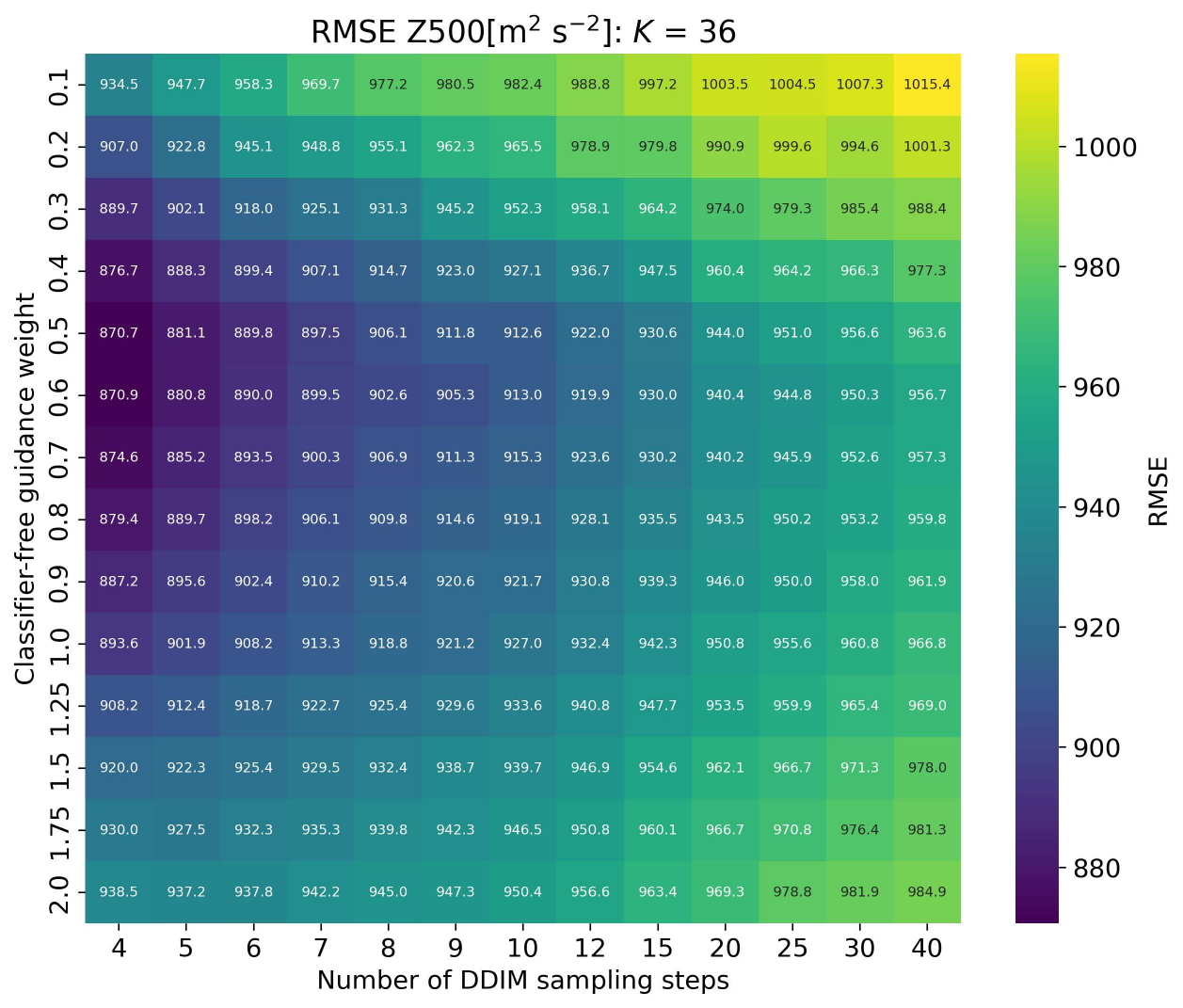}}
\caption{The effect of different classifier-free guidance weights and different number of DDIM sampling steps on RMSE Z500 for DM direct with $K=36$.}
\label{direct_cfg_k=36}
\end{center}
\end{figure}

We also demonstrate the impact of varying the classifier-free guidance weight on the Z500 RMSE for the DM direct model across different numbers of DDIM sampling steps, as illustrated in Figures \ref{direct_cfg_k=1} to \ref{direct_cfg_k=36}. We find that similar to the DM iterative model, as the lead time $K$ increases, a lower classifier-free guidance weight is needed to attain a similarly low RMSE. For long-range lead times that have $K$ beyond 20, a classifier-free guidance weight of around 0.5 is able to attain a lower RMSE for that specific long-range lead time $K$. This trend potentially suggests that sampling weather states for longer lead times, which inherently possess greater uncertainty, tends to align more closely with unconditional sampling, where the lead time $K$ is not specified. 

\section{Examples of Persistence and Weekly Climatology Guidance Diffusion}\label{persist_climo_examples}
\subsection{$t_0$ Configurations}\label{persist_climo_t0_config}
We provide detailed specification of $t_0$ for ``persist multi'' in \cref{Persistence_guidance} and ``Climo 144h refined'' in \cref{Climo_guidance} as the following:
\paragraph{``Persist $t_0$= multi''} begins with \( t_0 = 700 \) for lead times less than 24 hours. For each subsequent 24-hour period, \( t_0 \) is sequentially reduced: It first decreases by 100 for the 24 hours in the immediate future, and then it further decreases by 200 for each of the following 24-hour interval. This reduction continues until \( t_0 \) reaches 100 and remains at 100 until the lead time reaches 168 hours. Beyond this 168-hour lead time, \( t_0 \) is set to 50 for all remaining iterations.
\paragraph{``W\_Climo 144h $t_0 = $ multi''} initially employs persistence guidance with the same \( t_0 \) schedule as outlined in ``Persist $t_0$= multi''. Starting from the 144-hour lead time, it transitions to weekly climatology guidance, where \( t_0 \) starts at 250 and is gradually reduced to 100.

\subsection{Examples of Persistence Guidance}
\begin{figure}[ht]
\begin{center}
\centerline{\includegraphics[width=\columnwidth]{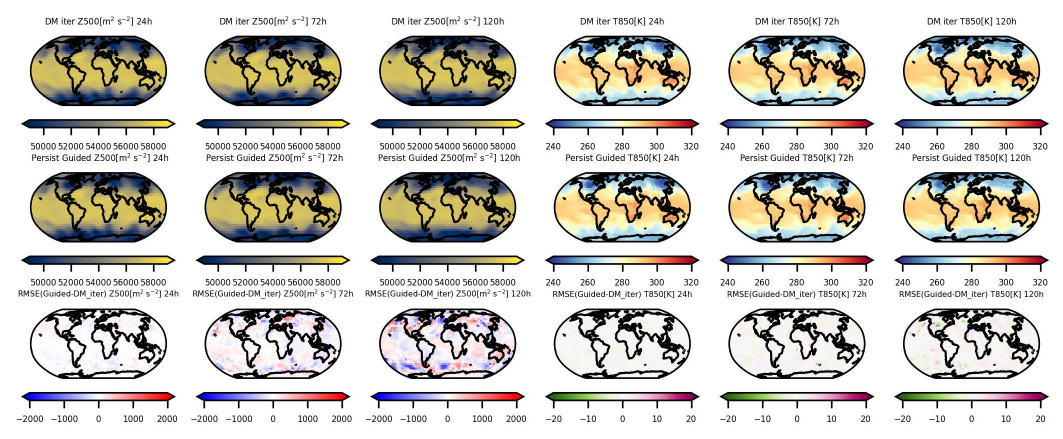}}
\caption{Similar to \cref{NWP_guidance_example}, but for persistence guidance represented by ``Persist $t_0$= multi'' in \cref{Persistence_guidance} for lead time from 24h to 120h. The top row represents the baseline DM iterative model for comparison.}
\label{persist_example_1}
\end{center}
\end{figure}

\begin{figure}[ht]
\begin{center}
\centerline{\includegraphics[width=\columnwidth]{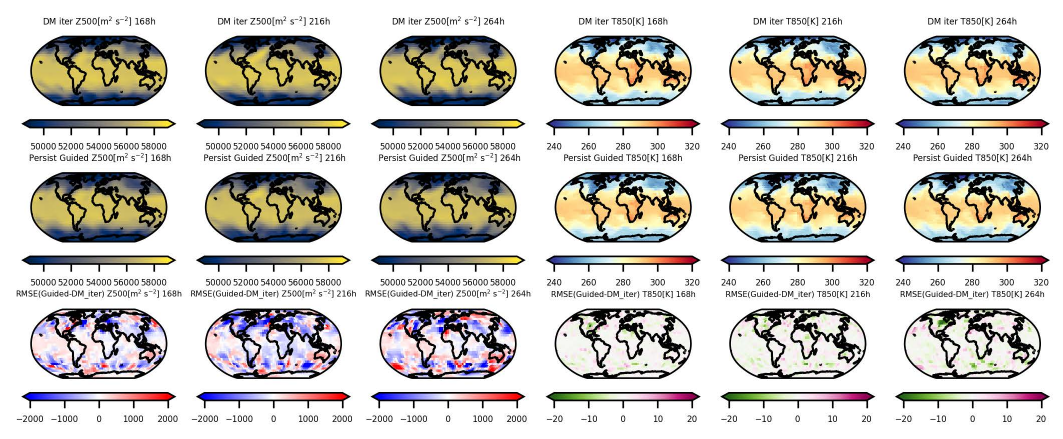}}
\caption{Similar to \cref{persist_example_1} but for lead time from 168h to 264h.}
\label{persist_example_2}
\end{center}
\end{figure}

We present examples of forecasts generated with persistence guidance in \cref{persist_example_1} and \cref{persist_example_2}.

\subsection{Examples of Climatology Guidance}\label{climo_examples}
\begin{figure}[ht]
\begin{center}
\centerline{\includegraphics[width=\columnwidth]{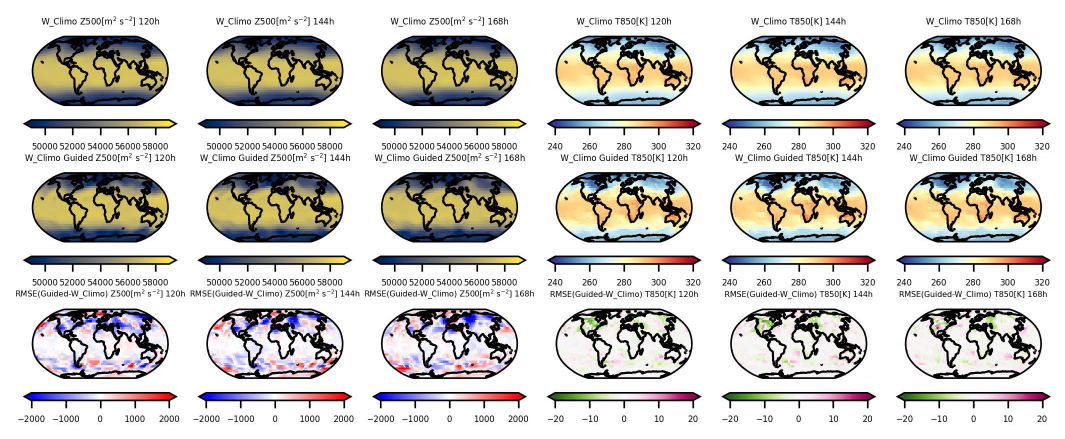}}
\caption{Similar to \cref{NWP_guidance_example}, but for weekly climatology guidance represented by ``W\_Climo 144h $t_0 = $ multi'' in \cref{Climo_guidance} for lead time from 24h to 120h. The top row represents the guidance used which is the weekly climatology forecast.}
\label{climo_example_1}
\end{center}
\end{figure}

\begin{figure}[ht]
\begin{center}
\centerline{\includegraphics[width=\columnwidth]{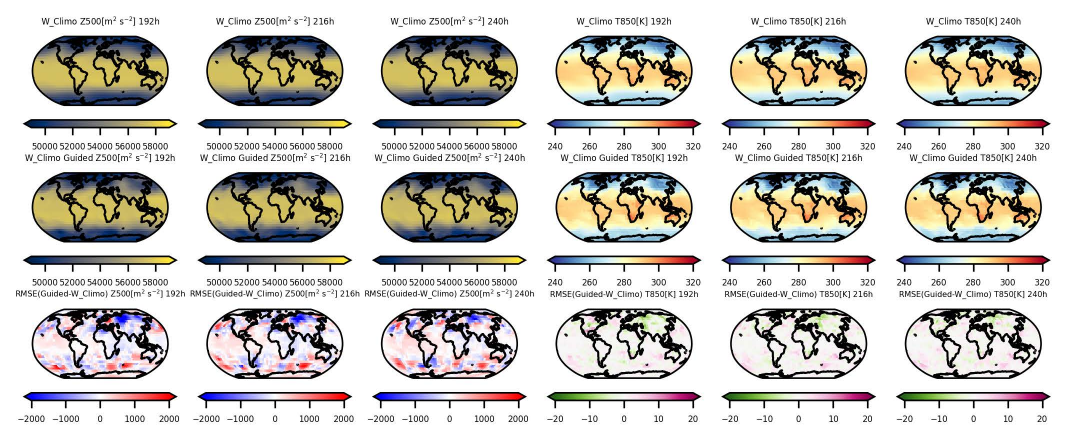}}
\caption{Similar to \cref{climo_example_1} but for lead time from 168h to 264h.}
\label{climo_example_2}
\end{center}
\end{figure}

We present examples of forecasts generated with climatology guidance in \cref{climo_example_1} and \cref{climo_example_2}.

\section{Failure Case Analysis and Limitations}\label{failure_analysis}

\begin{figure}[ht]
\begin{center}
\centerline{\includegraphics[width=\columnwidth]{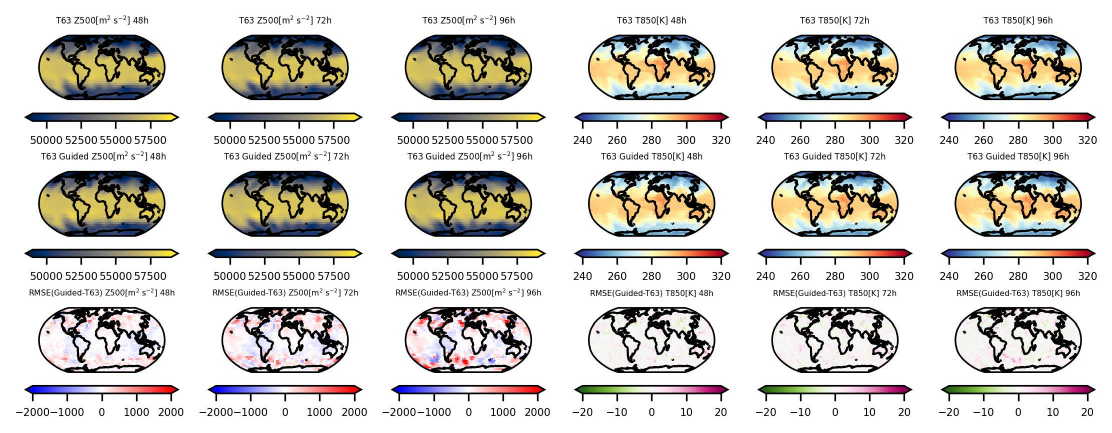}}
\caption{Similar to \cref{NWP_guidance_example}, but to demonstrate the limited performance when using IFS T63 guidance with initialize time of 2017-01-22 12UTC. We provide detailed failure case analysis in \cref{failure_analysis}.}
\label{NWP_failure_1_1}
\end{center}
\end{figure}

\begin{figure}[ht]
\begin{center}
\centerline{\includegraphics[width=\columnwidth]{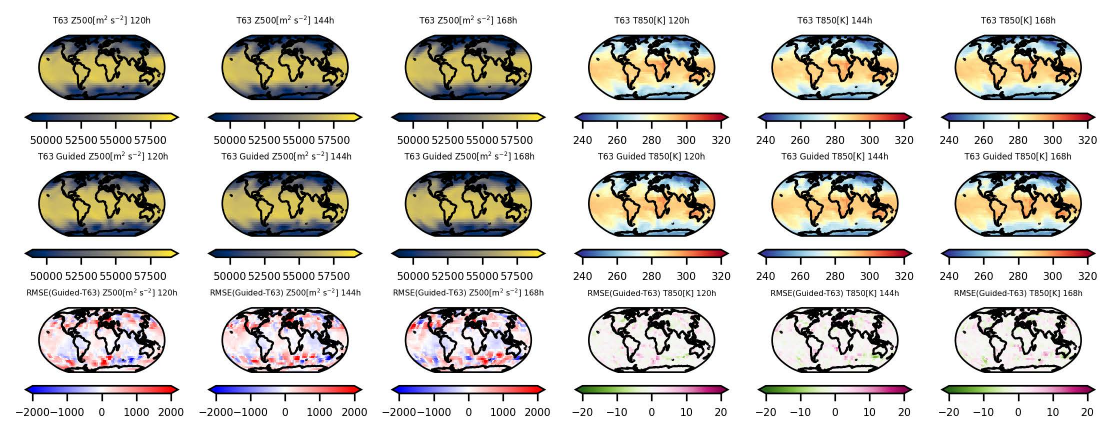}}
\caption{Similar to \cref{NWP_failure_1_1} but for lead time from 120h to 168h.}
\label{NWP_failure_1_2}
\end{center}
\end{figure}

\begin{figure}[ht]
\begin{center}
\centerline{\includegraphics[width=\columnwidth]{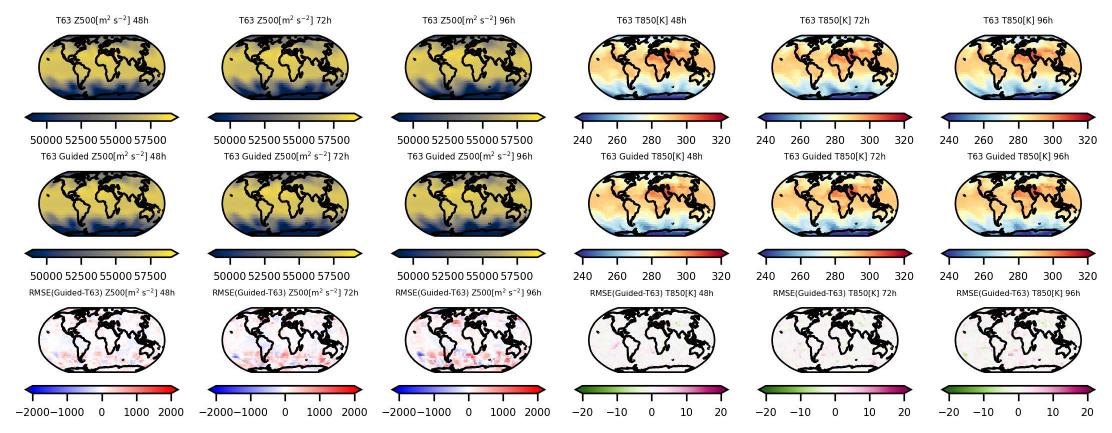}}
\caption{Similar to \cref{NWP_guidance_example}, but to demonstrate the limited performance when using IFS T63 guidance with initialize time of 2017-08-06 12UTC. We provide detailed failure case analysis in \cref{failure_analysis}.}
\label{NWP_failure_2_1}
\end{center}
\end{figure}

\begin{figure}[ht]
\begin{center}
\centerline{\includegraphics[width=\columnwidth]{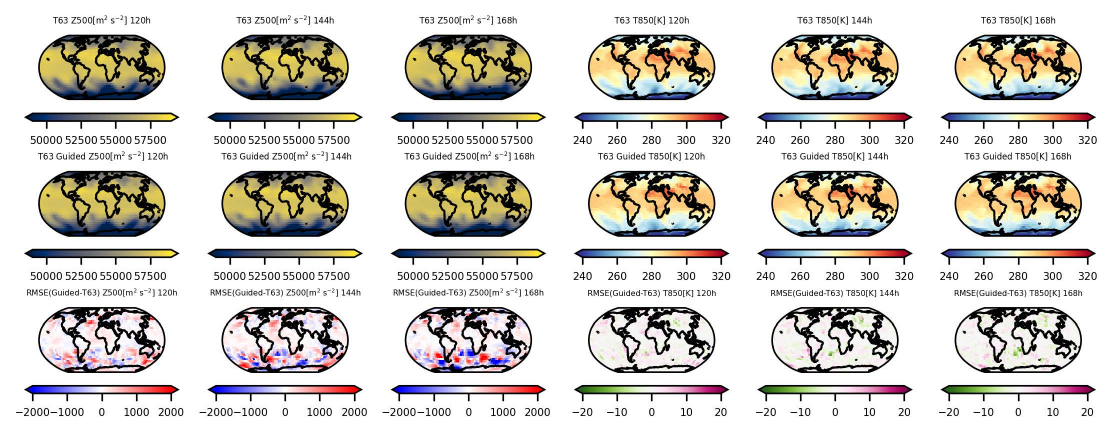}}
\caption{Similar to \cref{NWP_failure_2_1} but for lead time from 120h to 168h.}
\label{NWP_failure_2_2}
\end{center}
\end{figure}

In \cref{NWP_guidance}, we notice that using DM with a very strong NWP model as guidance may not improve the performance upon the NWP model itself. Examples using T63 as guidance similar to \cref{NWP_guidance_example} are provided for analysis in Figure \ref{NWP_failure_1_1}-\ref{NWP_failure_2_2}. We find that those guided predictions generated atmospheric fields, especially for T850 field, are less smooth in contrast to the more natural field in the top row of the images produced by T63 forecast.

Furthermore, unlike the application of image synthesis, besides the model's performance, there are other limitations or precautions when applying guidance specifically to weather forecasting. For example, the performance of the forecast for different weather fields could vary even using the same model. Thus when applying guidance towards output, $t_0$ should be selected carefully while considering the performance of all fields that the model produces. Moreover, climatology guidance tends to focus on improving the long-range forecast stability of the diffusion model. However, there are also other ways to calculate climatology with different temporal window sizes. Those different climatologies could require more testing when coupled with different $t_0$. In addition, persistence guidance is also field sensitive. One can imagine that if a field varies dramatically from one time to the next, persistence guidance when applied with small $t_0$ could dramatically reduce the forecast skill. We hypothesize that DM forecasting models that are designed to forecast large-scale weather conditions with smaller $\Delta K$ are more beneficial to apply persistence guidance and leave this analysis to future works.

